\documentclass[superscriptaddress]{revtex4-2}
\usepackage[monochrome]{color} 
\usepackage{amsmath,amssymb,amsfonts}
\usepackage{amsthm}
\numberwithin{equation}{section}
\theoremstyle{plain}%

\usepackage[normalem]{ulem} 
\usepackage{xcolor}

\allowdisplaybreaks
\begin{document}
\title{A coordinate-free approach to obtaining exact solutions in general relativity: The Newman-Unti-Tamburino  solution revisited}
\author{Emir Baysazan}
\affiliation{Department of Physics, Istanbul Technical University, 34469, Istanbul, Türkiye.}
\author{Ayşe Hümeyra Bilge}
\affiliation{Department of Industrial Engineering, Kadir Has University, 34083, Istanbul, Türkiye.}
\author{Tolga Birkandan}
\email{Contact author: E-mail: birkandant@itu.edu.tr}
\affiliation{Department of Physics, Istanbul Technical University, 34469, Istanbul, Türkiye.}
\author{Tekin Dereli}
\affiliation{Department of Physics, Koç University, 34450, Istanbul, Türkiye.}
\affiliation{Faculty of Aviation and Aeronautical Sciences, Özyeğin University, 34794, Istanbul, Türkiye.}
\date{\today}
\begin{abstract}
The Newman-Unti-Tamburino (NUT) solution is characterized as the unique Petrov Type $D$ vacuum metric such that the two double principal null directions form an integrable distribution. The uniqueness of the NUT is established by evaluating the integrability conditions of the Newman-Penrose equations up to  $SL(2,\mathbb{C})$  transformations, resulting in a coordinate-free characterization of the solution. 
\end{abstract}
\maketitle
\section{Introduction}

\textcolor{red}{The strategy for obtaining exact solutions of Einstein's equations is to start with a metric ansatz in a given coordinate frame, possibly with specific symmetry conditions and pre-determined sources along with their field equations.} In this scheme, Einstein's equations are second-order partial differential equations for the metric functions. The Newman-Penrose (NP) formalism, proposed in the early 1960s \cite{Newman:1961qr}, \textcolor{red}{provides} a useful tool for obtaining exact solutions; in particular, it is valuable for studying the asymptotic behavior of gravitational fields \cite{Stephani:2009exact}.

The Newman-Penrose formalism \textcolor{red}{is} a special case of the moving frame approach,\textcolor{red}{where} the connection is defined \textcolor{red}{using} the commutators of the frame as sections of the tangent bundle or the exterior derivatives of the frame as sections of the cotangent bundle. Then, Cartan's Structural Equations \textcolor{red}{define the curvature}, and the Bianchi Identities provide relations \textcolor{red}{between} the derivatives of the curvature components.

In the NP formalism, the moving frame consists of four null vector fields, \textcolor{red}{also known as} the ``null tetrad". The NP equations are first-order partial differential equations for the components of the connection and the curvature. Directional derivative operators are dual to the 1-forms of the moving frame.

The components of the connection are called ``spin coefficients''. The Newman-Penrose equations are \textcolor{red}{a way to calculate} the directional derivatives of the spin coefficients and define the curvature. The explicit expressions of the transformation of NP quantities under the rotations of the null tetrad, called ``gauge transformations," are given in \cite{Carmeli:1975wq}.

The commutation relations of the null tetrad, the NP equations, the Bianchi identities, and the gauge transformations constitute a coordinate-free framework for Einstein's equations. In the literature, the null tetrad is usually defined in terms of the metric, and coordinate transformations are used to express the final form of the exact solution in terms of local coordinates.   

In this work, we \textcolor{red}{ aim to characterize} exact solutions in a coordinate-free manner via the  integrability conditions of the NP equations. \textcolor{red}{Our approach begins} with \textcolor{red}{specific} algebraic and geometric ansatz expressed in terms of the NP quantities, \textcolor{red}{followed by an analysis of} the integrability of the resulting system of equations.  

Under \textcolor{red}{the given} geometric and algebraic assumptions, the NP system \textcolor{red}{is typically} overdetermined. A systematic study of the integrability conditions in the framework of the Riquier-Janet theory  \cite{schwarz1984} will result in either an inconsistent system or a system in ``involution"  for which the existence of a local solution is guaranteed. 

\textcolor{red}{In a system in involution, the remaining freedom is represented by functions} that are not determined by the integrability conditions. \textcolor{red}{Such a system can be regarded as defining an exact solution if this arbitrariness can be removed through an appropriate coordinate transformation. Although the procedure described above provides a coordinate-free method of solving the NP equations -without explicit reference to the metric-} it may be necessary to ``reconstruct" the metric in order to deal with the elimination of the coordinate freedom. 


In the present work, we \textcolor{red}{apply} this method to rederive the well-known Newman-Unti-Tamburino (NUT) solution, \cite{Newman:1963yy}, as a special type of Type $D$ vacuum metrics. 

\textcolor{red}{We assume} that the metric is vacuum and Petrov Type $D$; the geometric ansatze are that the repeated null directions of the Weyl tensor form an integrable distribution. In terms of NP quantities, this assumption leads to the condition $\tau+\bar{\pi}=0$, corresponding to the group (Type I) of Type $D$ vacuum metrics as classified by \cite{Kinnersley:1969zza}. Type $D$ metrics with repeated null directions forming an integrable distribution will be denoted as ``NUT background". Non-vacuum solutions on NUT backgrounds are being investigated by the present authors via the computation of integrability conditions. 

The Petrov classification \cite{Stephani:2009exact} is based on the structure of the Weyl tensor's eigenvector. In this classification, Type $D$ metrics are characterized by the existence of two principal null directions; hence, the spacetime naturally admits a $2+2$ decomposition. The Schwarzschild, NUT, Reissner-Nordström, and Kerr metrics are among the physically important Type $D$ vacuum solutions \cite{ahsan}.

In \cite{czapor1982orthogonal}, Czapor and McLenaghan studied all type $D$ solutions with a cosmological constant for the Newman-Penrose equations. They employed the commutation relations and examined various cases of $\mu,\rho,\tau$ and $\pi$. In \cite{debever1981orthogonal}, Debever showed that type $D$ vacuum and electrovacuum fields with a cosmological constant satisfy the Goldberg-Sachs theorem.

Our approach to \textcolor{red}{studying} exact solutions is \textcolor{red}{most closely related to} that of Edgar et al. \cite{edgar2009petrov}, \textcolor{red}{where} integrability conditions \textcolor{red}{form the core of} the solution procedure. In \cite{edgar2009petrov}, Petrov Type $D$ vacuum solutions are investigated using integrability within the GHP formalism. The notion of ``complete tables" is introduced as those quantities for which derivatives in all directions are determined. The paper presents a comprehensive review of Type $D$ metrics, including two identities among spin coefficients, namely  $\mu\bar\rho=\bar\mu\rho$ and $\tau\bar\tau=\pi\bar\pi$.

The structure of the paper is as follows. Section \ref{sec2} \textcolor{red}{introduces} the NP equations as an overdetermined system. Section \ref{sec3} characterizes Type $D$ metrics using the Petrov classification invariants and the Jordan forms of the matrix $Q$. In Section \ref{appb}, we discuss the relevant $SL(2,\mathbb{C})$ transformations for Type $D$ metrics. In Section \ref{sec5} we derive the integrability conditions for Petrov Type $D$ vacuum metrics under the assumptions $\rho \neq 0$ and $\rho \neq \bar{\rho}$, and compute the dimension of the corresponding symmetry algebra, which allows us to characterize the NUT solution. In Section \ref{sec6}, we prove that the NUT solution is the unique vacuum Type $D$ metric whose principal null directions form an integrable distribution, showing that $\bar{\tau}+\pi=0$ implies $\tau=0$ and that under these assumptions the NP equations admit a unique solution up to gauge and coordinate transformations. Section \ref{sec7} determines the symmetry algebra of Type $D$ vacuum metrics and characterizes the NUT solution through its Killing vectors. In Section \ref{sec4}, we discuss vacuum Type $D$ spacetimes under the condition $\rho = \bar{\rho}$, which corresponds to non-twisting null congruences. In Section \ref{sec9}, \textcolor{red}{we calculate the commutation relations for the metric components} and show that the remaining freedom in the solution of the involutive system corresponds to a diffeomorphism of the underlying manifold. The appendices contain the NUT metric in local coordinates, the required mathematical definitions, and the list of NP equations for general Type $D$ metrics.

\section{Overdetermined systems of partial differential equations}\label{sec2} 

A system of partial differential equations is said to be in normal form when each equation determines a derivative of a dependent variable with respect to the same independent variable. In particular, the number of equations should be equal to the number of unknowns. \textcolor{red}{According to the} well-known Cauchy-Kovalevskaya (CK) theorem, if a system of partial differential equations is in normal form, then there exists a unique solution in a neighborhood of the initial (non-characteristic) surface. \textcolor{red}{For some} physical applications of the CK theorem, see \cite{maia,izumi2013analysis,ita2008proposal,klainerman1999local}.

When symmetry conditions or other geometric constraints are imposed, Einstein's equations may no longer \textcolor{red}{appear} in normal form, and the number of equations can exceed the number of unknowns.

Systems in which the number of equations exceeds the number of variables are called overdetermined. The analogue of the CK theorem for the existence of solutions is the Riquier-Janet theorem for overdetermined systems \cite{schwarz1984}. This theorem states that if the system is in ``involution,'' then an analytic solution exists, with the free parameters determined by the structure of the system in involution.

This solution method \textcolor{red}{arises from} the theory of overdetermined systems \textcolor{red}{developed} by Riquier and Janet \textcolor{red}{at the turn of the 20th century}. It relies on the computing of integrability conditions, \textcolor{red}{which determine whether the system is inconsistent -thus admitting no solution- or can be brought into involution, in which case the Riquier-Janet existence theorem guarantees the existence of a local analytic solution.} For instance, the Riemann-Lanczos problem can be investigated using the Riquier-Janet theorem \cite{lanczos1949lagrangian, bampi1983third, dolan2003exterior}.
 
\textcolor{red}{To illustrate the role of integrability conditions, we consider the example discussed} in reference \cite{brans1967computer}, where only the trivial solution exists. In this system, the independent variables are $\{x,y,z\}$, and the dependent variables are $\{u,v,w\}$. If the system of equations were of the form 
\begin{equation*}
 u_x=-vx,\quad v_x=uy, \quad w_z=u+\epsilon,   
\end{equation*}
then the first two equations would form a normal system to which the CK theorem could be applied, and $w$ could then be solved. However, if we introduce a third equation for $u$ and $v$, as shown below 
\begin{equation*}
u_x=-vx,\quad u_y=-wz,\quad v_x=uy, \quad w_z=u+\epsilon,
\end{equation*}
the system's structure changes, affecting the applicability of the CK theorem.

The $y$ derivative of the first equation should be consistent with the $x$ derivative of the second equation, since mixed partial derivatives commute.
The addition of this new equation leads to an overdetermined system, and in  \cite{brans1967computer}, it is shown that no nontrivial solution exists.

The key ingredient of the study of integrability conditions is to define a total order on the set of partial derivatives of the dependent variables and solve each equation with respect to the derivatives with the highest rank with respect to this order \cite{schwarz1984}. 

The Newman-Penrose formalism is a version of the Cartan moving frame method for a 4-dimensional spacetime metric and a lightlike moving frame (null tetrad) \cite{Newman:1961qr}. \textcolor{red}{A comprehensive} bibliography on this formalism can be found in \cite{Stephani2003}. The Newman-Penrose equations are first-order equations defined by directional derivatives. 

The method of ordering partial derivatives and computing integrability conditions can be applied to the Newman-Penrose system, provided that the commutation relations are used appropriately.

In the Newman-Penrose system, \textcolor{red}{the derivatives are ordered as follows}. In the spirit of first-order formulation, we \textcolor{red}{adopt} a scheme in which the Bianchi identities are considered as differential equations for the Weyl and Ricci spinors. The unknown quantities are $\Psi_i \ ($i=0,\dots,4$),$ $\Phi_{ij} \  ($i,j=0,1,2$) $, and $\Lambda $. The ranks of $\Psi_i,\Phi_{ij}$ and $\Lambda$ are chosen to be higher than those of the spin coefficients.

Regarding the derivatives, and motivated by the decomposition of the NP equations into radial and non-radial equations—commonly used in the study of exact solutions—we rank the directional derivatives as below. For any function $\phi$,  
\begin{equation}
D\phi>\Delta\phi>\{\delta\phi,\overline{\delta}\phi\}.
\end{equation}
The ordering between $\delta\phi$ and $\overline{\delta}\phi$ depends on which one is primarily defined by the NP equations. For example, $\delta\rho>\bar\delta\rho$, while $\bar\delta\mu> \delta\mu$. Furthermore, if $\delta\phi>\bar\delta\phi$, then $\bar\delta\bar\phi>\delta\bar\phi$, and vice versa. 

\section{Canonical forms of Type $D$ metrics}\label{sec3}

\textcolor{red}{Petrov type $D$ spacetimes are characterized by the Jordan–Segre decomposition of the complex, symmetric, traceless $3 \times 3$ Weyl matrix $Q$ defined in the NP formalism. The multiplicities of the eigenvalues of the $Q$ matrix determine the Petrov type \cite{Petrov:2000bs}. This eigenvalue structure allows the canonical form of type $D$ metrics—in which only $\Psi_2$ remains nonzero—to be obtained via suitable $SL(2,\mathbb{C})$ transformations \cite{Carmeli:1975wq}. A detailed canonical classification of type $D$ metrics and their invariant structures was developed by \cite{Ferrando2002}. A short review of the canonical forms of type $D$ solutions along the lines of previous literature \cite{Stephani:2009exact,Baak2023,McIntosh_1994,dinverno71} would be helpful.}

In the Petrov classification, Type $D$ metrics are classified as algebraically special metrics characterized by the vanishing of the invariants $K$ and $N$ as defined in \cite{Stephani:2009exact}, for $\Psi_4\ne 0$. Following the general conventions adopted in the literature, if the null vector $l$ is the repeated principal null direction, then $\Psi_0=\Psi_1=0$, and Type $D$ metrics are characterized by the equality $3\Psi_2\Psi_4=2\Psi_3^2$, and by using  $SL(2,\mathbb{C})$ rotations it is possible to set $\Psi_3=\Psi_4=0$.  In addition, for the vacuum case, $\Psi_2$  turns out to be the only surviving curvature component.  

Petrov classification is also characterized in terms of the Jordan forms of the  $3\times 3$ complex, symmetric traceless matrix  $Q$, defined as below \cite{Stephani:2009exact}.
\begin{equation*}
Q=
\begin{pmatrix}
\Psi_2-\frac{1}{2}(\Psi_0+\Psi_4) & \frac{i}{2}(\Psi_4-\Psi_0)        & \Psi_1-\Psi_3\\
\frac{i}{2}(\Psi_4-\Psi_0)        & \Psi_2+\frac{1}{2}(\Psi_0+\Psi_4) & i(\Psi_1+\Psi_3)\\
\Psi_1-\Psi_3                     & i(\Psi_1+\Psi_3)                  & -2\Psi_2
\end{pmatrix}. 
\end{equation*}
For Type $D$ metrics, $Q$ has a simple eigenvalue $k$ and a repeated eigenvalue $\frac{1}{2}k$, with $k\ne 0$, hence its minimal polynomial $m(t)=(t+\textstyle{\frac{1}{2}}k I)(t-kI)=0.$ Evaluating $m(t)$ with $t=Q$, the matrix equation 
\begin{equation*}
    E=(Q+\textstyle{\frac{1}{2}}k I)(Q-k I)=0,
\end{equation*}    
gives the following conditions
\begin{align*}
 E_{11} &:\quad  k^2 -\textstyle {\frac{4}{3}}\Psi_0\Psi_4 +\textstyle{\frac{16}{3}}\Psi_1\Psi_3 -4 \Psi_2^2=0,  \\
 E_{12} &:\quad k \Psi_0-4 \Psi_0 \Psi_2 +4 \Psi_1^2=0, \\
E_{13} &:\quad  k \Psi_1 -2 \Psi_0\Psi_3 +2 \Psi_1 \Psi_2=0,\\
E_{22}&:\quad  k \Psi_4 -4 \Psi_2\Psi_4 +4 \Psi_3^2 =0,\\
E_{23}&:\quad  k \Psi_3 -2 \Psi_1\Psi_4 +2 \Psi_2 \Psi_3=0,\\
E_{33}&:\quad  k \Psi_2 -\textstyle{\frac{2}{3}}\Psi_0\Psi_4 
            -\textstyle{\frac{4}{3}}\Psi_1\Psi_3+2 \Psi_2^2=0. 
\end{align*}
Solutions of these equations can be characterized as follows.
\begin{itemize}
    \item Case (i): $\Psi_1=0$, $\Psi_3=0$, $\Psi_0=0$, $\Psi_4=0$. In this case $\Psi_2\neq 0$ and $k=-2\Psi_2$.
     \item Case (ii): $\Psi_1=0$, $\Psi_3=0$, $\Psi_0\ne0$.  As $\Psi_0\ne 0$, $E_{12}$ implies that $k=4\Psi_2$, and substituting this in $E_{33}$, it can be seen that $\Psi_4\ne0$. $\Psi_0\Psi_4=9\Psi_2^2$.
   \item Case (iii): $\Psi_1=0$, $\Psi_3\ne 0$. In this case $k=-2\Psi_2$. The equation $E_{13}$ is reduced to  $\Psi_0\Psi_3=0$, hence $\Psi_0=0$ and $3\Psi_2\Psi_4=2\Psi_3^2$.
 \item Case (iv): $\Psi_1\ne 0$, $\Psi_3=0$. This is the analogue of case (iii) and results in $k=-2\Psi_2$, $\Psi_4=0$, and $3\Psi_2\Psi_0=2\Psi_1^2$.
\item Case (v): $\Psi_1\ne0$,$\Psi_3\ne 0$. In this case, $k=4\left(\Psi_2 -\textstyle{\frac{\Psi_1^2}{\Psi_0}}\right)$ and $\Psi_3$ and $\Psi_4$ are expressed as 
\begin{equation*}
    \Psi_3=\textstyle{\frac{\Psi_1}{\Psi_0^2}} \left(3\Psi_0\Psi_2 -2\Psi_1^2\right),\quad
\Psi_4=\textstyle{\frac{1}{\Psi_0^3}} 
\left(9\Psi_0^2\Psi_2 -12\Psi_0 \Psi_1^2\Psi_2 +4\Psi_1^4\right). 
\end{equation*} 
\end{itemize}
We use the following sequence of $SL(2,\mathbb{C})$ transformations. We start by setting $\Psi_0=0$ by a Type C rotation as given in \cite{Carmeli:1975wq}. Then  equation $E_{12}$ implies that $\Psi_1=0$ and Type A rotations for $\Psi_i$, $i=2,3,4$ are
\begin{equation*}
\Psi_2'=\Psi_2,\quad \Psi_3'=3 z \Psi_2+\Psi_3,\quad
\Psi_4'=6 z^2\Psi_2+4z\Psi_3+\Psi_4. 
\end{equation*}
We use this rotation to set $\Psi_3'=0$, and the condition $3\Psi_2'\Psi_4'=2\Psi_3'^2$  implies that $\Psi_4'=0$. Thus, up to $SL(2,\mathbb{C})$ rotations, $\Psi_2$ turns out to be the only surviving component of the Weyl tensor. 
 
\section{$SL(2,\mathbb{C})$ transformations}\label{appb} 

Explicit expressions of the transformations of the spin coefficients and the curvature components \textcolor{red}{under $\mathrm{SL}(2,\mathbb{C})$ transformations} are given in \cite{Carmeli:1975wq}. 
\textcolor{red}{These transformations are used to simplify  the system and  fix the tetrad to obtain a complete classification possibly after coordinate transformations, as for example in 
 \cite{Kinnersley:1969zza} for the complete classification of Type $D$ vacuum metrics.}
 
The $\mathrm{SL}(2,\mathbb{C})$  transformations are arranged in three groups; 
(i) null rotations about $l^\mu$;
(ii)  boost in the $l^\mu$-$n^\mu$ plane and spatial rotation in the $m^\mu$-$\bar{m}^\mu$ plane;
(iii) null rotations about $n^\mu$.
\textcolor{red}{In the literature, a common gauge consists of choosing 
 $\kappa = \epsilon = \pi = 0$.  When 
 $l_{\mu}$ is hypersurface orthogonal, meaning that it is proportional to a gradient field, then $\bar{\rho} = \rho$ and for such gradient congruences, the condition $\tau = \bar{\alpha} + \beta$ also holds  \cite{Newman:1961qr}.}

\textcolor{red}{ For Type $D$ metrics with $\Psi_i=0$ for $i\ne 2$,  $\mathrm{SL}(2,\mathbb{C})$ transformation consisting of rotations about the null vectors $l^\mu$ and $n^\mu$ are not allowed, as they would violate the conditions $\Psi_i=0$, for $i\ne 2$. Hence, only boosts in the $l^\mu$-$n^\mu$ plane and spatial rotations in the $m^\mu$-$\bar m^\mu$ plane, denoted as ``Type B transformations" are allowed. For a vacuum Type $D$ metric the Bianchi identities imply that $\kappa = \sigma = \lambda = \nu = 0$. 
 In Proposition IV.1, we will show that under these conditions one can set $\epsilon=0$  and $\tau = \bar{\alpha} + \beta$, by Type B transformations even in the case $\rho-\bar\rho\ne 0$, provided that $\tau+\bar\pi=0$. Then in Proposition IV.2, we will prove that for a vacuum Type $D$ metric, the condition $\tau+\bar\pi=0$ implies $\tau=0$, hence the condition $\tau+\bar\pi=0$  corresponds to the ``Case 1, $\pi^o=\tau^o=0$" in  \cite{Kinnersley:1969zza}. Finally, in Proposition IV.3, we will show that the imaginary part of $\gamma$ can also be fixed by Type B transformations. }

\vskip 0.2 cm \noindent
{\bf Proposition IV.1.} {\it Let $\kappa=0$. $SL(2,\mathbb{C})$ transformations consisting of boost in the $l^\mu$-$n^\mu$ plane and spatial rotation in the $m^\mu$-$\bar{m}^\mu$ plane can be used to set 
\begin{equation}
\epsilon=0,\quad \quad  \tau=\bar\alpha+\beta,
\end{equation} provided that $(\tau+\bar\pi)(\rho-\bar\rho)=0$.}
\vskip0.2cm\noindent
{\it Proof.}
The null tetrad transforms as
\begin{equation*}
 l'_\mu=Al_\mu,\quad m'_\mu=e^{i\theta}m_\mu,\quad n'_\mu=A^{-1}n_\mu, 
 \end{equation*}
where  $z=A^{1/2}e^{i\theta/2}$,  $A$ and $\theta$ are real. 
\begin{align*}
&\rho'=A\rho,                        &      &\mu' = A^{-1}\mu,  \tag{IV.2a}\\
&\pi' = e^{-i\theta}\pi,             &      &\tau'= e^{ i\theta}\tau, \tag{IV.2b}\\
&\epsilon'= A(\epsilon-zDz^{-1}),    &      &\gamma'  = A^{-1}(\gamma -z\Delta z^{-1}),\tag{IV.2c} \\
&\alpha'  = e^{-i\theta}(\alpha -z\bar{\delta} z^{-1}),  &    & \beta'  = e^{i\theta}(\beta -z\delta z^{-1}).\tag{IV.2d}
\end{align*}
Choosing $Dz=-z\epsilon$ leads to  $\epsilon'=0$ and further transformations should be compatible with $Dz=0$. The term $\tau-\bar\alpha-\beta$ transforms as  
\begin{eqnarray*}
\tau'-\bar{\alpha}'-\beta'&=& e^{i\theta}(\tau-\bar{\alpha}-\beta +\bar{z}\delta \bar{z}^{-1}+z\delta z^{-1}), \\
&=& e^{i\theta}\left(\tau-\bar{\alpha}-\beta 
-\frac{\delta \bar{z}}{\bar{z}}
-\frac{\delta z}{z}\right), \\
&=& e^{i\theta}\left(\tau-\bar{\alpha}-\beta 
-\frac{\delta {A}}{A}\right). 
\end{eqnarray*}
Thus, $\bar{\alpha}'+\beta'$ can be set to zero, provided that 
$DA=0$ and $\delta A=(\tau-\bar{\alpha}-\beta)A$ are compatible.
For $\kappa=0$, $\epsilon=0$ and $DA=0$, the  commutator of $D$ and $\delta$ given by Eqn.(\ref{B.1b}) 
applied to $A$ gives
\begin{eqnarray*}
D \delta A&=& \sigma \bar{\delta} A+\bar{\rho}\delta A,\\
D (\tau-\bar{\alpha}-\beta)A&=& 
\sigma (\bar\tau-{\alpha}-\bar\beta)A
+\bar{\rho}(\tau-\bar{\alpha}-\beta)A,\\
D (\tau-\bar{\alpha}-\beta)&=& 
\sigma (\bar\tau-{\alpha}-\bar\beta)
+\bar{\rho}(\tau-\bar{\alpha}-\beta).
\end{eqnarray*}
Using the NP equations Eqn.(\ref{B.6a},\ref{B.7a},\ref{B.7b}) for $D\tau$, $D\bar\alpha$ and $D\beta$, for $\kappa=0$ and $\epsilon=0$ are
substituted in the  commutation relation to result in 
$(\tau+\bar\pi)(\rho-\bar\rho)=0$,
and the proof is completed.\qed
\vskip 0.2cm

\textcolor{red}{We will now prove that the condition $\tau+\bar\pi=0$ implies $\tau=\pi=0$ for a vacuum Type $D$ spacetime. We will prove this implication under slightly more general conditions.}

\textcolor{red}{\vskip 0.2cm\noindent{\bf Proposition IV.2.}
{\it Let $M$ be a  vacuum metric of Petrov Type D with $\Psi_i=0$ for $i\ne 2$ and $\Psi_2-\overline\Psi_2\ne 0$. Then $\tau+\bar\pi=0$ implies $\tau=\pi=0$.}}
\vskip 0.1cm \noindent
\textcolor{red}{{\it Proof.}
For vacuum metrics with $\Psi_i=0$ for $i\ne 2$, the Bianchi identities imply that $\kappa=\nu=\sigma=\lambda=0$. If $\tau+\bar\pi=0$,  by Proposition IV.1, one can set $\epsilon=0$ and $\tau=\bar\alpha+\beta$. NP equations for $\tau$  give
\begin{equation*}
D\tau=\Phi_{01}+\Psi_1,\quad \Delta\tau=(\gamma-\bar\gamma)\tau+\Phi_{12}+\bar\Psi_3, \quad D\gamma=\tau\pi+\Psi_2+\Phi_{11}-\Lambda,
\end{equation*}
and their commutation relation is
$(\Delta D-D\Delta)\tau=(\gamma+\bar\gamma)D\tau$.
Substituting the derivatives of $\tau$ and $\gamma$ we obtain,
\begin{equation*}
(\Delta-2\gamma)(\Phi_{01}+\Psi_1)-D(\Phi_{12}+\bar\Psi_3)=(\Psi_2-\bar\Psi_2)\tau.
\end{equation*}
Thus, as $\Phi_{01}+\Psi_1=0$ and $\Phi_{12}+ \bar \Psi_3=0$ and $\Psi_2\ne \bar\Psi_2$, it follows that $\tau=0$.\qed
}
\vskip 0.2cm

\textcolor{red}{The condition $\tau=\pi=0$ can be expressed as a geometric constraint, as follows. Let $(M,g)$ be a metric of Petrov type $D$. The principal null directions of the Weyl tensor naturally indicate a local $2+2$ decomposition of the manifold $M$, expressed as $M=B\times F$. In the NP formalism, we may choose the tetrad such that $(l,n)$ are the principal null directions of the Weyl tensor, and $(l,n)$ and $(m,\bar m)$  form local bases of the cotangent bundles of $B$ and $F$, respectively. A Type $D$ spacetime is characterized as ``twisting" provided that its principal null directions are twisting. 
If $(M,g)$ is a vacuum Type $D$ spacetime \textcolor{red}{with $\Psi_2\ne 0$ only}, the spin coefficients satisfy $\kappa=\nu=\sigma=\lambda=0$. These conditions are expressed in terms of the commutation relations, as follows,}
\begin{align}
[D,\delta]&\in {\rm Span}(D,\delta),\quad (\kappa=\sigma=0), \tag{V.5a}\label{43}\\
[\Delta,\delta]&\in {\rm Span}(\Delta,\delta),\quad (\nu=\lambda=0),\tag{V.5b}\label{44} \\
[D, \Delta]&\in {\rm Span}(D,\Delta),\quad (\tau+\bar\pi=0), \tag{V.5c}\label{45}\\
[D, \delta]&\in {\rm Span}(\delta),\quad
(\kappa=\sigma=\bar\alpha+\beta-\bar\pi=0), \label{46} \tag{V.5d}\\
[\Delta, \delta]&\in {\rm Span}(\delta),\quad
(\nu=\lambda=\bar\alpha+\beta-\tau=0). \tag{V.5e} \label{47}
\end{align}
\textcolor{red}{In terms of these conditions, Proposition IV.2 can be reformulated as follows. For a vacuum spacetime, conditions  (\ref{43}) and (\ref{44}), (\ref{45}) implies (\ref{46}) and (\ref{47}), that lead to $\tau-\bar\pi=0$, hence to $\tau=0$, after appropriate gauge transformations. This result does not hold for the non-twisting case. There are solutions with $\tau+\bar\pi=0$, with $\tau\ne 0$, $\pi\ne 0$. We can thus reformulate Proposition IV.2 as below.}
\textcolor{red}{
\vskip 0.2cm\noindent{\bf Proposition IV.2. (rephrased)}
{\it Let $(M,g)$ be a vacuum metric of Petrov Type D. If the principal null directions of the Weyl tensor, denoted by $(D,\Delta)$, are twisting $(\rho \neq \bar{\rho}, \, \mu \neq \bar{\mu})$ and define an integrable distribution $(\tau + \bar{\pi} = 0)$, then the orthogonal complement of this distribution is preserved under the action of the Lie derivative operator along the principal null directions.}}

\vskip 0.2cm
\textcolor{red}{ We will now prove that the imaginary part of $\gamma$ can also be fixed by an $SL(2,\mathbb{C})$ transformation.}

\vskip 0.2cm\noindent{\bf Proposition IV.3.}
{\it Let $(M,g)$ be a vacuum Type $D$ spacetime with $\Psi_2\ne 0$ only and with  $\kappa=\sigma=0$, $\epsilon=0$, $\tau+\bar\pi=0$ and $\tau=\bar\alpha+\beta$. Then $SL(2,\mathbb{C})$ transformations consisting of boost in the $l^\mu$-$n^\mu$ plane and spatial rotation in the $m^\mu$-$\bar{m}^\mu$ plane can be used to set 
\begin{equation}
\gamma'-\bar{\gamma}'=\textstyle{\frac{1}{2}}
\left(
\frac{\Psi_2}{\rho'}-\frac{\overline{\Psi}_2}{\bar{\rho}'}\right). \tag{IV.3}
\label{IV.3}
\end{equation}}
\vskip 0.1cm\noindent
{\it Proof.}
The imaginary part of $\gamma$ transforms as 
\begin{eqnarray*}
\gamma'-\bar{\gamma}'&=& A^{-1}
\left(  \gamma -\bar{\gamma}
-z\Delta z^{-1}+\bar{z}\Delta \bar{z}^{-1}
\right), \\
&=& A^{-1}
\left(  \gamma -\bar{\gamma}
+\frac{\Delta z}{z}-\frac{\Delta \bar{z}}{\bar{z}}
\right),\\
&=& A^{-1}
\left(  \gamma -\bar{\gamma}
+i\Delta\theta\right). 
\end{eqnarray*}
We look for the consistency of the equation 
\begin{eqnarray*}
A^{-1}
\left(  \gamma -\bar{\gamma}
+i\Delta\theta\right)&=& \textstyle{\frac{1}{2}}
\left(
\frac{\Psi_2}{\rho'}-\frac{\bar{\Psi}_2}{\bar{\rho}'}\right), \\
A^{-1}
\left(  \gamma -\bar{\gamma}
+i\Delta\theta\right)&=& \textstyle{\frac{1}{2}}
\left(
\frac{\Psi_2}{A\rho}-\frac{\bar{\Psi}_2}{A\bar{\rho}}\right), \\
\left( \gamma -\bar{\gamma}
+i\Delta\theta\right)&=& \textstyle{\frac{1}{2}}
\left(
\frac{\Psi_2}{\rho}-\frac{\bar{\Psi}_2}{\bar{\rho}}\right).
\end{eqnarray*}
Since $D\theta=0$, applying $D$ to the last equation, we obtain,
\begin{eqnarray*}
D\gamma -D\bar{\gamma}
+iD\Delta\theta&=& \textstyle{\frac{1}{2}}
\left(
\frac{D\Psi_2}{\rho}-\frac{\Psi_2 D\rho}{\rho^2}
-\frac{D\bar{\Psi}_2}{\bar{\rho}}
+\frac{\bar{\Psi}_2 D\bar{\rho}}{\bar{\rho}^2}
\right), \\
\Psi_2 -\bar{\Psi}_2
+iD\Delta\theta&=& \textstyle{\frac{1}{2}}
\left(
\frac{3\rho\Psi_2}{\rho}
-\frac{\Psi_2 \rho^2}{\rho^2}
-\frac{3\bar{\rho}\bar{\Psi}_2}{\bar{\rho}}
+\frac{\bar{\Psi}_2 \bar{\rho}^2}{\bar{\rho}^2}
\right), \\
i D\Delta\theta&=& 0, \\
\Delta D\theta-(\gamma+\bar{\gamma})D\theta&=& 0, 
\end{eqnarray*}
and the last equation is identically satisfied since $D\theta=0$.
$\hfill\qed$
\vskip 0.2cm
\section{ Type $D$ VACUUM METRICS} \label{sec5}

In \cite{Kinnersley:1969zza}, Kinnersley obtained \textcolor{red}{the complete class of} Type $D$ vacuum metrics using the Newman-Penrose formalism. Several cases yield solutions which can coarsely be divided into $\rho=0$, $\rho\neq 0$, but $\rho=\bar\rho$, and  $\rho\neq  0$, $\rho\neq \bar\rho$ subcases. The NUT metric  \cite{Newman:1963yy}, whose local coordinate \textcolor{red}{representation} is \textcolor{red}{given} in Appendix \ref{appnut}, corresponds to the case $\rho \neq 0$, $\pi_0=\tau_0=0$. 

In this section, we derive the integrability conditions for Petrov Type $D$ vacuum metrics under the assumptions $\rho \neq 0$ and $\rho \neq \bar{\rho}$. We then compute the dimension of the corresponding symmetry algebra, 
which allows us to characterize the NUT solution by means of this dimension.

Appendix B presents the commutators of directional derivative operators, the exterior derivatives of the basis $1$-forms and the Killing equations for the most general case. The NP equations are given for Type $D$ vacuum metrics, i.e, for 
$  \Phi_{ij}=\Lambda=0$.
For Petrov Type $D$ metrics, as discussed in Section \ref{sec3}, the null tetrad can be chosen such that only $\Psi_2$ remains non-vanishing, i.e. $\Psi_0 = \Psi_1 = \Psi_3 = \Psi_4 = 0$. Under these conditions, the Bianchi identities that involve $\Psi_2$ algebraically reduce to the following set of equations:
\begin{equation}
3 \kappa\Psi_2=0,\quad 3 \sigma \Psi_2=0, \quad 3 \lambda \Psi_2=0, \quad 3 \nu \Psi_2=0.
\end{equation} 
Since $\Psi_2$ is nonzero, \textcolor{red}{it follows that} for all Type $D$ vacuum metrics with repeated principal null directions $l^\mu$, $n^\mu$ we have
\begin{equation}
\kappa=\sigma=\nu=\lambda=0.\label{TypeD}
\end{equation}
In addition, we use $SL(2,\mathbb{C})$ rotations of Type B to set
$\epsilon=0$. 
\vskip 0.3cm\noindent
{\bf Integrability of NP equations for Type D Vacuum metrics:}
The Bianchi identities for Type $D$ metrics \textcolor{red}{take the form}
\begin{equation}
D\Psi_2=3 \rho \Psi_2,\quad\quad
\Delta\Psi_2=-3\mu \Psi_2,\quad\quad
\bar{\delta}\Psi_2=-3\pi\Psi_2,\quad\quad \delta\Psi_2=3 \tau\Psi_2.
\end{equation}
\textcolor{red}{Applying the commutation relations of the directional derivative operators to $\Psi_2$ and its complex conjugate $\bar \Psi_2$ yields:}
\begin{align}
[D,\Delta]\Psi_2 \to \quad &\bar\delta\tau=-\delta\pi+(\alpha-\bar\beta)\tau+(\bar\alpha-\beta)\pi-\mu\bar\rho+\bar\mu\rho, \tag{V.4a} \label{V.4a}\\
[D,\Delta]{\bar\Psi}_2\to \quad&\delta\bar\tau=-\bar\delta\bar\pi+(\bar \alpha-\beta)\bar\tau+(\alpha-\bar\beta)\bar\pi-\bar\mu\rho+\mu \bar\rho,\tag{V.4b} \label{V.4b}\\
[\Delta,\delta]\Psi_2\to\quad
&\Delta\tau=-\delta\mu-( \bar\alpha+\beta)\mu+\tau(\gamma-\bar\gamma),\tag{V.4c} \label{V.4c}\\
[\Delta,\delta]{\bar\Psi}_2\to\quad
&\Delta\bar\tau=-\bar\delta\bar\mu-(\alpha+\bar\beta)\bar\mu+\bar\tau(\bar\gamma-\gamma),\tag{V.4d} \label{V.4d}\\
[D,\bar\delta]\Psi_2\to\quad
&D\pi=-\bar\delta\rho+(\alpha+\bar\beta)\rho,\tag{V.4e} \label{V.4e}\\
[D,\bar\delta]\Psi_2\to\quad
&D\bar\pi=-\delta\bar\rho+(\bar\alpha+\beta)\bar\rho. \tag{V.4f} \label{V.4f}
\end{align}
The system of NP equations, supplemented by the additional relations above, is \textcolor{red}{then} solved with respect to the following sets of derivatives: 
\begin{align*}
    &\{D\rho,\Delta\rho,\delta\rho\},&                   &\{D\bar\rho,\Delta\bar\rho,\bar\delta\bar\rho\},&
    &\{D\mu, \Delta\mu,\bar\delta\mu\},&                  &\{D\bar\mu, \Delta\bar\mu ,\delta\bar\mu     \},&\\
    &\{D\tau,\Delta\tau,\delta\tau,\bar\delta\tau\},&    &\{D\bar\tau,\Delta\bar\tau,\delta\bar\tau,\bar\delta\bar\tau\},&
    &\{D\pi,\Delta\pi,\bar\delta\pi\},&                  &\{D\bar\pi,\Delta\bar\pi,\delta\bar\pi\},&                     \\
    &\{D\alpha,\Delta\alpha,\delta\alpha\}, &            &\{D\bar\alpha,\Delta\bar\alpha,\bar\delta\bar\alpha\},&     
    &\{D\beta,\Delta\beta\},&                            &\{D\bar\beta,\Delta\bar\beta\},               \\
    &\{D\gamma\},&                                       &\{D\bar\gamma\}.&                           \end{align*}
For all variables except $\tau$ and $\bar\tau$, commutation relations of $D$ and $\Delta$  derivatives are identically satisfied. The compatibility of $D$ and $\Delta$ derivatives of $\tau$ and $\bar\tau$ gives the following second-order equation for $\delta\delta\pi$  and its complex conjugate:
\begin{equation}
\delta\delta\pi=
\delta(\bar\alpha -\beta)\pi
+\delta\pi \ (\bar\alpha -\beta+2 \tau)
+\delta\mu \ (\rho-2\bar\rho)
-\mu \delta\bar\rho   
+(\bar\alpha+\beta)(\mu\rho-\mu\bar\rho)-2(\bar\alpha+\beta)\pi\tau+\tau(4 \beta \pi + 2\mu \bar\rho -\Psi_2 -\bar\Psi_2). \tag{V.5} \label{V.5}
\end{equation}
The compatibility of $\delta\rho$, $\bar{\delta}\mu$, and $\delta\alpha$, together with their complex conjugates, with the $D$ and $\Delta$ derivatives impose no additional constraints. However, from the compatibility of  $\delta\tau$ and $\bar{\delta}\pi$ with their respective $D$ and $\Delta$ derivatives, we obtain second order equations for $\delta\delta\mu$ and $\bar{\delta}\bar{\delta}\rho$, along with their conjugates.
\begin{align}
\delta\delta\mu&=
-\delta(\bar\alpha+\beta)\ \mu
-\delta\mu \ (3\bar\alpha+\beta-3\tau)
-2\bar\alpha (\bar\alpha+\beta)\mu
+3(\bar\alpha+\beta)\mu\tau,  \tag{V.6a} \label{V.6a}\\
\bar\delta\bar\delta\rho&=
\bar\delta(\alpha+\bar\beta)\ \rho
+\bar\delta\rho\ (\alpha+3\bar\beta-3\pi)
-2\bar\beta (\alpha+\bar\beta)\rho 
+3(\alpha+\bar\beta)\rho\pi. \tag{V.6b} \label{V.6b}
\end{align}
\textcolor{red}{Verification of the compatibility between} these second-order derivatives (and their conjugates) with the previously defined three first-order derivatives, which together yield $18$ integrability conditions. After performing the necessary substitutions and algebraic manipulations, one obtains the following well-known identities:
\begin{equation}
      \rho \bar{\mu} = \bar{\rho} \mu,   \quad
\tau \bar{\tau} = \pi \bar{\pi}. \tag{V.7} \label{V.7}
\end{equation}
Additionally, the first-order derivatives of $\rho$ and $\mu$ together with their conjugates are derived:
\begin{align}
\bar\delta\rho&=
(\alpha+\bar\beta-2\pi-\bar\tau)\rho+\pi \bar\rho, \tag{V.8a} \label{V.8a}\\
\delta\mu&=(-\bar\alpha-\beta+2\tau+\bar\pi)\mu-\tau \bar\mu. \tag{V.8b} \label{V.8b}
\end{align}
Second-order equations for $\rho$ and $\mu$ are identically satisfied, but  the compatibility of these equations with previous ones gives four first-order equations for $\delta\pi$ and $\bar\delta\bar\pi$ of the form
\begin{align}
&\bar\tau\ \delta\pi+\pi \ \bar\delta\bar\pi=P,\quad\quad 
\bar\pi \ \delta\pi+\tau\ \bar\delta\bar\pi=Q, \tag{V.9a} \label{V.9a}\\
&\bar\rho\ \delta\pi-\rho\ \bar\delta\bar\pi=R,\quad\quad 
\bar\mu \ \delta\pi+\mu \ \bar\delta\bar\pi=S,\tag{V.9b} \label{V.9b}
\end{align}
where $P$, $Q$, $R$ and $S$  are polynomial expressions of the spin coefficients and in $\Psi_2$.  
This system can be solved for $\delta\pi$ and $\bar\delta\bar\pi$ provided that 
$\varphi=\bar\tau\rho+\pi\bar\rho$ is nonzero. This can be seen by showing that $D\varphi=0$ is incompatible with $\varphi=0$ for $\rho-\bar\rho\ne 0$ or $\tau\ne 0$.
Finally, it has been checked that all four derivatives of the algebraic relations are satisfied, and after substituting all identities, we have the following integrable system:
\begin{align}
D\Psi_2&=3\rho\Psi_2, \tag{V.10a} \label{V.10a}\\
\Delta\Psi_2&=-3\mu\Psi_2, \tag{V.10b} \label{V.10b}\\
\delta\Psi_2&=3\tau\Psi_2, \tag{V.10c} \label{V.10c}\\
\bar\delta\Psi_2&=-3\pi\Psi_2. \tag{V.10d} \label{V.10d} \\ \nonumber \\
D\rho&=\rho^2, \tag{V.11a} \label{V.11a}\\
\Delta\rho&=-\delta\pi+(\bar\alpha-\beta)\pi+(\gamma+\bar\gamma-\bar\mu)\rho -\tau\bar\tau-\Psi_2, \tag{V.11b} \label{V.11b}\\
\delta\rho&=(\bar\alpha+\beta)\rho+\tau(\rho-\bar\rho), \tag{V.11c} \label{V.11c}\\
\bar\delta\rho&=(\alpha+\bar\beta)\rho-(\bar\tau+\pi)\rho-\pi(\rho-\bar\rho). \tag{V.11d} \label{V.11d} \\ \nonumber \\
D\mu &=	\delta(\pi) - \bar \alpha \pi + \beta \pi + \mu \bar \rho + \pi \bar \pi + \Psi_2, \tag{V.12a} \label{V.12a} \\
  \Delta\mu &=	- \mu(\gamma + \bar \gamma + \mu), \tag{V.12b} \label{V.12b} \\
  \delta\mu &= - \bar \alpha  \mu - \beta \mu + \mu \bar \pi + 2 \mu \tau - \bar \mu \tau, \tag{V.12c} \label{V.12c}\\
  \bar \delta \mu &= - \alpha \mu - \bar \beta \mu - \mu \pi + \bar \mu \pi	. \tag{V.12d} \label{V.12d} \\  \nonumber\\
D\tau &=\rho(\bar \pi + \tau), \tag{V.13a} \label{V.13a} \\
\Delta \tau &=	\gamma \tau - \bar \gamma \tau - \mu \bar \pi - 2 \mu \tau + \bar \mu \tau, \tag{V.13b} \label{V.13b} \\			
\delta\tau &=	\tau(- \bar \alpha + \beta + \tau), \tag{V.13c} \label{V.13c} \\			
\bar \delta\tau &=	- \delta(\pi) + \alpha \tau + \bar \alpha \pi - \beta \pi - \bar \beta \tau - \mu \bar \rho + \bar \mu \rho. \tag{V.13d} \label{V.13d} \\ \nonumber \\
D\pi& =	2 \pi \rho - \pi \bar \rho + \rho \bar \tau, \tag{V.14a} \label{V.14a} 	\\		
\Delta \pi &=	- \gamma \pi + \bar \gamma  \pi - \mu \pi - \mu \bar \tau, \tag{V.14b} \label{V.14b}\\ 
\bar\delta\pi &= \pi(- \alpha + \bar \beta - \pi). \tag{V.14c} \label{V.14c} \\ \nonumber \\ 		
D \alpha &=	\rho(\alpha + \pi),  \tag{V.15a} \label{V.15a}\\		\Delta\alpha &=	\bar \delta(\gamma) + \alpha \bar \gamma - \alpha \bar \mu + \bar \beta \gamma - \gamma \bar \tau, \tag{V.15b} \label{V.15b}\\			
\delta\alpha &=	\bar \delta(\beta) + \alpha \bar \alpha - 2 \alpha \beta + \beta \bar \beta + \gamma \rho - \gamma \bar \rho + \mu \rho - \Psi_2. \tag{V.15c} \label{V.15c} \\ \nonumber \\	
D\beta &=	\beta \bar \rho,  \tag{V.16a} \label{V.16a}\\		
\Delta\beta &= \delta(\gamma) + \bar \alpha \gamma + 2 \beta \gamma - \beta \bar \gamma - \beta \mu - \gamma \tau - \mu \tau. \tag{V.16b} \label{V.16b}\\ \nonumber \\	
D \gamma &=	\alpha \bar \pi + \alpha \tau + \beta \pi + \beta \bar \tau + \pi \tau + \Psi_2,	 \tag{V.17} \label{V.17}
\end{align}
where $\delta(\pi)$ is
\begin{align}
\delta(\pi)&= (\bar \alpha \pi^2 \bar \rho + \bar \alpha \pi \rho \bar \tau - \beta \pi^2 \bar \rho - \beta \pi \rho \bar \tau + 2 \mu \pi \rho \bar \rho - 2 \mu \pi \bar \rho^2 - \pi \Psi_2 \bar \rho +\pi \bar \Psi_2 \rho 
+ 2 \pi \rho \tau \bar \tau - \pi \bar \rho \tau \bar \tau + \rho \tau \bar \tau^2)/(\pi \bar \rho + \rho \bar \tau). \tag{V.18} \label{V.18} 
\end{align}
From these equations, it can be seen that the freedom in the metric consists of the following derivatives, 
$$\{\bar\delta\alpha\},\quad 
\{\delta\beta,\bar\delta\beta\}, 
\{\Delta\gamma,\delta\gamma,\bar\delta\gamma\},\quad
\{\Delta\bar\gamma,\delta\bar\gamma,\bar\delta\bar\gamma\}.$$
This leads to $9$  free parameters in the solution. Together with the spin coefficients $\rho$, $\bar\rho$, $\mu$, $\tau$, $\bar\tau$, $\pi$, $\alpha$, $\bar\alpha$, $\beta$, $\bar\beta$ and $\gamma$, $\bar\gamma$ that are not   algebraically determined, there are all together  $20$ free parameters in the general Type $D$ vacuum solution.  Possible gauge transformations and the characterization of the subcases will be presented elsewhere.

\section{Characterization of the NUT Solution} \label{sec6}
In this section, we prove that the NUT solution is uniquely determined as the vacuum type $D$ metric where the principal null directions of the Weyl tensor form an integrable distribution. This actually amounts to proving that the condition $\bar\tau+\pi=0$  implies $\tau=0$, and showing that under these assumptions the NP equations have a unique solution up to gauge and coordinate transformations. \textcolor{red}{We will start our calculations by following Kinnersley \cite{Kinnersley:1969zza} closely. Then we will show that the equation (VI.16) we derived implies that the $2$-dimensional space-like submanifold with co-tangent fields $m$ and $\overline{m}$  has constant curvature.}


The conditions $\tau=\pi=0$ single the  NUT solution out among Type $D$ vacuum metrics, as stated in \cite{{Kinnersley:1969zza}}. Here,  we prove that for twisting solutions, characterized by $\Psi_2\ne \bar\Psi_2$ (hence $\rho\ne \bar\rho$), $\tau=\pi=0$ is a consequence of the geometric assumption $\tau+\bar\pi=0$.  On the other hand, for $\rho=\bar\rho$ there exist solutions satisfying $\tau+\bar\pi=0$, while $\tau\neq 0$, as shown in Section \ref{sec4}. Moreover, in Section \ref{appb} it was proved that if either $\tau+\bar\pi=0$ or $\rho=\bar\rho$, then $SL(2,\mathbb{C})$ transformations of Type B can be used to set 
\begin{equation}
\epsilon=0,\quad \tau=\overline{\alpha}+\beta, \quad \gamma-\bar\gamma=
\textstyle{\frac{1}{2}}\left(
{\frac{\Psi_2}{\rho}-\frac{\bar{\Psi}_2}{\bar\rho}}\right).
\label{gauge}
\end{equation}

\vskip 0.2cm\noindent
{\bf The NP system and algebraic eliminations:}
The Newman-Penrose system is grouped as follows:
\begin{align}
&D\Psi_2=3\rho \Psi_2, &  &\Delta\Psi_2= -3\mu \Psi_2,                                        &    &\delta \Psi_2=\overline{\delta} \Psi_2=0, \tag{VI.2a}      \label{psi2} \\
&D\rho= \rho^2,        &  &\Delta\rho= \rho(\gamma+\overline{\gamma}- \overline{\mu})-\Psi_2, &    &\delta\rho= 0,          &   &                          \tag{VI.2b}  \label{rho} \\
&D\mu= \overline{\rho}\mu+\Psi_2, &  &\Delta\mu= -\mu^2 -\mu(\gamma+\overline{\gamma}),       &    &\overline{\delta}\mu=0, &   &                        \tag{VI.2c}   \label{mu} \\
&D\alpha= \rho\alpha,  &  &\delta \alpha +\overline{\delta}\overline{\alpha}= \rho\mu+4\alpha\overline{\alpha}+\gamma(\rho-\overline{\rho})-\Psi_2, & &    \tag{VI.2d} \label{alpha1} \\ 
&\Delta\alpha-\overline\delta{\gamma}= \alpha(\overline{\gamma}-\gamma-\overline{\mu}),       &                 &\delta\gamma+\Delta\overline{\alpha}= -\overline{\alpha}(\mu+\overline{\gamma}-\gamma),      &             &D\gamma=\Psi_2.         \tag{VI.2e}\label{gamma}
\end{align}
Note that the left-hand side of the second equality in equation (\ref{alpha1}) is real, thus the imaginary part of its right-hand side has to be zero, namely,
\begin{equation}    
   (\gamma+\overline\gamma)(\rho-\overline\rho)+\mu\rho-\bar{\mu}\bar{\rho}-(\Psi_2-\overline\Psi_2)=0 \tag{VI.3}.\label{rgamma}
\end{equation}
It follows that $(\gamma+\overline\gamma)$ is determined provided that $(\rho-\overline\rho)$ is non-zero. The case $\rho=\overline{\rho}$ will be studied in Section \ref{sec4}. Thus, assuming $(\rho-\overline\rho)\ne 0$, and taking into account the gauge condition given by Eq.(\ref{gauge}), $\gamma$  is completely determined. Although we need to substitute $\gamma$ into Eqs.(\ref{rho}-\ref{gamma}) and check for consistency, we postpone this at the moment and investigate the integrability of the system for $\Psi_2$.
\vskip 0.5cm\noindent
{\bf The NP system and integrability conditions:}
The commutators of the differential operators are 
\begin{align}  
&\Delta D-D\Delta=(\gamma+\overline\gamma)D,&
&\delta D-D\delta=-\overline\rho \delta,\cr
&\delta \Delta-\Delta \delta=(\mu-\gamma+\overline\gamma)\delta,&
&\overline\delta\delta-\delta\overline\delta=(\overline\mu-\mu)D+(\overline\rho-\rho)\Delta
                 +2\alpha \delta-2\overline\alpha\overline\delta. \nonumber 
\end{align}
\vspace{0.2cm}
\noindent \textbf{System of equations for $\Psi_2$:}
Substituting  $D\Psi_2$ and $\Delta \Psi_2$  in the  commutator of $D$ and $\Delta$ to $\Psi_2$, we obtain
\begin{equation*}
3 \Psi_2 \left(\Delta \rho \right)+3 \rho \left(\Delta \Psi_2 \right)+3 \Psi_2 \left(D \mu\right)+ 3 \mu \left(D \Psi_2\right)=\left(\gamma+ \overline{\gamma}\right)D \Psi_2. 
\end{equation*}
Then, after substituting $D\mu$, $\Delta\rho$, $D\Psi_2$, and $\Delta\Psi_2$, we obtain the algebraic relation
\begin{equation}
    \mu \bar{\rho} - \bar{\mu} \rho = 0.  \tag{VI.4}
\end{equation}
Next,  the commutator of $D$ and $\bar{\delta}$ applied to $\Psi_2$ gives
\begin{equation*}
    3 \Psi_2 \left(\bar{\delta} \rho\right) + 3 \rho \left(\bar{\delta} \Psi_2\right) = 0. 
\end{equation*}
As $\overline{\delta}\Psi_2=0$, we obtain $\overline{\delta}\rho=0$. Similarly, the commutator of $\Delta$  and $\delta$ applied to $\Psi_2$ gives $\delta\mu=0$.
Thus all derivatives of $\rho$ and $\mu $ are completely determined, as 
\begin{equation}
\bar \delta \rho=0,\quad\quad \delta \mu=0.  \tag{VI.5}
\end{equation}
Finally,  the commutation relation of $\delta$ and $\bar{\delta}$ is satisfied provided that $\mu\bar{\rho}=\bar{\mu}\rho$.
At this stage, all derivatives of $\Psi_2$ are consistently defined and the only freedom in $\Psi_2$ is an arbitrary constant.
Actually, we can ``solve" $\Psi_2$ as
\begin{equation}
\Psi_2=\rho^3\Psi_{20},  \tag{VI.6}
\end{equation}
and it can be checked that $D\Psi_{20}=\Delta\Psi_{20}=\delta\Psi_{20}=\bar{\delta}\Psi_{20}=0$, thus  $\Psi_{20}$ is a complex constant.

\vspace{0.2cm}\noindent \textbf{System of equations for $\rho$ and $\mu$:}
Recall that all derivatives of $\rho$ have been completely determined. Applying the commutation relation of $\delta$ and $\bar{\delta}$ to $\rho$ yields
\begin{equation*}
    (\bar\mu-\mu)D\rho+(\bar\rho-\rho)\Delta\rho=0.
\end{equation*}
This leads to the algebraic relation
\begin{equation*}
2\mu\rho\overline\rho-2\mu\overline{\rho}^2-\Psi_2\overline\rho +\overline\Psi_2\rho=0. 
\end{equation*}
By substituting $\mu \bar{\rho}=\bar{\mu}\rho$, we obtain the following symmetric form
\begin{equation*}
2\rho\overline\rho(\mu-\overline\mu)-\Psi_2\overline\rho +\overline\Psi_2\rho=0, 
\end{equation*}
hence the imaginary part of $\mu$ is determined as
\begin{equation}
\mu-\bar{\mu}=\frac{1}{2}\left(\frac{\Psi_2}{\rho}-\frac{\bar{\Psi}_2}{\bar{\rho}}\right).  \tag{VI.7}
\end{equation}
The remaining integrability conditions are identically satisfied and the only freedom in $\rho$ is also an arbitrary constant as in the case of $\Psi_2$. By  direct computations, we obtain the relations
\begin{equation}
D\left(\frac{1}{\rho}+\frac{1}{\bar{\rho}}\right)=-2,\quad
D\left(\frac{1}{\rho}-\frac{1}{\bar{\rho}}\right)=0.\quad  \tag{VI.8}
\end{equation}
Thus, up to a coordinate transformation,  $\rho$ takes the well-known form
$\rho=-(r+i\rho^\circ)^{-1}$,
where $\rho^\circ$ is a real constant.

Finally, replacing $\bar{\mu}=\mu(\bar{\rho}/\rho)$ in the expression for the imaginary part of $\mu$, we obtain the definition of $\mu$ as
\begin{equation}
    \mu=\frac{1}{2}\left[1-\frac{\bar{\rho}}{\rho}\right]^{-1}
\left(\frac{\Psi_2}{\rho}-\frac{\bar{\Psi}_2}{\bar{\rho}}\right).  \tag{VI.9}
\end{equation}
It has been checked that the differential equations for $\mu$ are satisfied. 

\vspace{0.2cm}\noindent \textbf{System of equations and gauge transformation for $\gamma$:}
In Section \ref{appb}, we show that a gauge transformation can be used to fix the imaginary part of $\gamma$. Together with the algebraic relation for its real part and substituting the previously derived expression for $\mu$, we obtain
\begin{equation}
  \gamma=\frac{1}{2}\frac{\Psi_2}{\rho}.  \tag{VI.10} 
\end{equation}
It has been checked that all derivatives are again identically satisfied.

\vspace{0.5cm}\noindent \textbf{System of equations for $\alpha$ and the freedom in the metric:}
The equation $D\alpha=\rho\alpha$ can be solved as
\begin{equation}
\alpha=\rho \alpha_0,\quad D\alpha_0=0.  \tag{VI.11}
\end{equation}
Substituting $\alpha$ and $\gamma$ in the equation for $\Delta\alpha$, it is easy to see that $\Delta\alpha_0=0$. 

After relevant substitutions, the non-radial equation for $\alpha$ becomes
\begin{equation}
\delta\alpha+\overline\delta\overline\alpha
-4\alpha\overline\alpha=-\frac{1}{2}\frac{1}{\rho-\overline{\rho}} \frac{1}{\rho\bar{\rho}}
\left[\bar{\Psi}_2\rho^3-\Psi_2\bar{\rho}^3\right].   \tag{VI.12} 
\end{equation}
Using  $\delta=\overline\rho \delta_0$, with 
$\delta_0D-D\delta_0=0$,   
we obtain an equation for $\alpha_0$, namely
\begin{equation}
\delta_0\alpha_0+\overline\delta_0\overline\alpha_0
-4\alpha_0\overline\alpha_0=-\frac{1}{2}
\frac{\rho\bar{\rho}}{\rho-\overline{\rho}} 
\left[\bar{\Psi}_{20}-\Psi_{20}\right],   \tag{VI.14}
\end{equation}
which is equivalently written as
\begin{equation}
    \delta_0\alpha_0+\overline\delta_0\overline\alpha_0
-4\alpha_0\overline\alpha_0=-\frac{1}{2}
\left[  \frac{1}{\bar{\rho}} -\frac{1}{\rho}\right]^{-1}
\left[\bar{\Psi}_{20}-\Psi_{20}\right],  \tag{VI.15}
\end{equation}
that finally gives
\begin{equation}
\delta_0\alpha_0+\overline\delta_0\overline\alpha_0
-4\alpha_0\overline\alpha_0=-\frac{1}{4i\rho_0}
\left[\bar{\Psi}_{20}-\Psi_{20}\right].  \tag{VI.16}
\end{equation}

We now show that this equation implies that the $2$-dimensional space-like submanifold with co-tangent fields $m$ and $\overline{m}$  has constant curvature.
\vskip 0.2cm\noindent{\bf Curvature of the submanifold with cotangent frame $\{m,\overline{m}\}$:} Let $M$ be a submanifold of a manifold $\tilde{M}$. The connection $\tilde\nabla$ on $\tilde{M}$  induces a connection $\nabla$ on $M$ and the corresponding curvatures can be defined accordingly.  
Let $e=\{e_1,e_2\}$ be an orthonormal co-frame for the cotangent bundle of a $2$-dimensional (Riemannian) submanifold. The connection and the curvature of $M$ are given  by $SO(2)$-valued (skew-symmetric) $2\times 2$ matrices $A$ and $R$, respectively, defined by
$de=Ae$, $R=dA-A\wedge A$. In component form, 
\begin{equation*}
    de_1=\omega \wedge e_2,\quad de_2=-\omega \wedge e_1,
\end{equation*}
where $\omega=\omega_1 e_1+\omega_2 e_2$.
If we work with  a complex frame $\{m,\overline{m}\}$, with $m=e_1+ie_2$, then it can be seen that 
\begin{equation}\label{SB:1}
dm=i\omega\wedge m,\quad d\overline{m}=i\omega\wedge \overline{m}.  \tag{VI.17}
\end{equation}
The exterior derivative of $m$, restricted to the span of  
$\{m,\overline m\}$, is
\begin{equation}\label{SB:2}
dm_{|M}=(\beta-\overline{\alpha})m\wedge\overline{m}.  \tag{VI.18}
\end{equation}
Comparing \eqref{SB:1} and \eqref{SB:2}, one can see that 
\begin{equation}\label{SB:3}
\omega=-i(\alpha-\overline{\beta})m+i(\overline{\alpha}-\beta)
\overline{m}.  \tag{VI.19}
\end{equation}
The curvature  $2$-form $\Omega$ is defined by 
$\Omega=d\omega-\omega\wedge \omega$, but as $\omega\wedge \omega=0$, it is just given by $\Omega=d\omega$. Taking the exterior derivatives of \eqref{SB:3}
\begin{equation}\label{SB:4}
    \Omega=\left[
    -id(\alpha-\overline{\beta})\wedge m
    -i(\alpha-\overline{\beta})\wedge dm
    +id(\overline{\alpha}-\beta)\wedge \overline{m}
    +i(\overline{\alpha}-\beta) \wedge d\overline{m}\right]_{|M}.  \tag{VI.20}
\end{equation}
The action of the exterior derivative $d$ on functions is given by 
$d=\Delta l + D n -\overline{\delta} m-\delta \overline{m}$.
Applying this to  \eqref{SB:4} and projecting to the submanifold, we obtain
\begin{equation}\label{SB:5}
 \Omega=-i\left[\delta(\alpha-\overline{\beta})
 +\overline \delta(\overline \alpha-{\beta})
 -2(\alpha-\overline\beta)(\overline\alpha-\beta)\right]m\wedge\overline{m},  \tag{VI.21}
\end{equation}
which reduces to the expression below for $\overline\alpha+\beta=0$.
\begin{equation}\label{SB:6}
\Omega=-2i \left[  \delta\alpha+\overline\delta\overline\alpha
-4\alpha\overline\alpha\right]m\wedge\overline{m}.  \tag{VI.22}
\end{equation}
Eliminating the $\rho$  dependency and using the equation
\begin{equation}
\delta_0\alpha_0+\overline\delta_0\overline\alpha_0
-4\alpha_0\overline\alpha_0=-\frac{1}{4i\rho_0}
\left[\bar{\Psi}_{20}-\Psi_{20}\right],  \tag{VI.23}
\end{equation}
we see that the submanifold $M$, with cotangent frame 
$\{m,\overline{m}\}$ has constant curvature.

\section{The Symmetry Algebra} \label{sec7}

In this section, we determine the symmetry algebra of Type $D$ vacuum metrics and characterize the NUT solution in terms of the symmetry algebra. \textcolor{red}{The Killing vector is parametrized as  
$X=A  D + B \Delta + C \delta +\bar C \bar \delta$.}
For the Type $D$ metrics, where $\kappa=\nu=\sigma=\lambda=0$, the Killing equations (\ref{B.10a}-\ref{B.10j}) \textcolor{red}{reduce to},
\begin{align*}
&DA + \left(\Delta-\gamma - \bar{\gamma}\right)B - \left(\tau- \bar{\pi}\right)C - \left(\bar{\tau}-\pi\right)\bar{C}=0, \\
&DB =0, \\
& \left(D+\rho\right)C    - \left(\bar{\delta}-\alpha-\bar{\beta}-\pi\right)B  =0, \nonumber \\
&  \left(D+\bar{\rho}\right)\bar{C} - \left(\delta-\bar{\alpha}-\beta-\pi\right)B  =0,  \\
&\left(\Delta+\gamma +\bar{\gamma}\right)A =0, \\
& \left(\Delta+\gamma-\bar{\gamma}-\bar{\mu}\right)C -\left(\bar{\delta}+\bar{\tau}+\alpha+\bar{\beta}\right)A =0, \nonumber \\
& \left(\Delta+\bar{\gamma}-\gamma-\mu\right)\bar{C}-\left(\delta+\tau+\bar{\alpha}+\beta\right)A =0, \nonumber \\
& \left(\bar{\delta}+\alpha-\bar{\beta}\right)C =0, \nonumber \\
&\left(\delta+\bar{\alpha}-\beta\right)\bar{C} =0, \nonumber \\
 & \left(\delta-\bar{\alpha}+\beta\right)C + \left(\bar{\delta}-\alpha+\bar{\beta}\right)\bar{C}-A\left(\rho+\bar{\rho}\right) + B \left(\mu + \bar{\mu}\right)=0.
\end{align*}
These equations \textcolor{red}{can be} ``solved" with respect to the following sets of derivatives:
 \begin{equation}\label{SA0}
\{DA,\Delta A\},\quad \{DB\},\quad 
\{DC,\Delta C,\bar \delta C,\delta C\},\quad 
\{D\bar C,\Delta \bar C, \delta \bar C\},
\end{equation}
They yield $10$ integrability conditions \textcolor{red}{determining} the second derivatives:
 \begin{equation}\label{SA1}
\Delta\Delta B,\quad
\delta \Delta B, \quad
\bar\delta\Delta B,\quad
\delta\delta B, \quad
\bar\delta\bar\delta B,\quad
\bar\delta \delta B,\quad 
\delta \delta A,\quad
\bar\delta\delta A,\quad
\bar\delta \bar\delta A,\quad
\bar\delta\bar\delta \bar C. 
\end{equation}
The integrability conditions of the equations that define the derivatives above are as follows. As the $D$ derivatives of $A$, $B$, $C$ and $\bar C$ are assigned, $D$ derivatives of all equations defining the derivatives in Eqn.(\ref{SA1}) should be added in the set of integrability conditions. Then, as $\Delta A$,  $\Delta C$ and $\Delta\bar C$ are assigned, $\Delta$ derivatives of the last $4$ derivatives in Eqn. (\ref{SA1}), and finally $\delta$  derivative of $\bar C$  are added to the set of integrability conditions, resulting in $15$ third-order equations, determining the following derivatives.
\begin{align}
&D\Delta\Delta B,\quad   D\delta \Delta B,\quad   D\bar\delta\Delta B,\quad    
D\delta\delta B,\quad   D\bar\delta\bar\delta B,\quad   D\bar\delta \delta B,\nonumber\\
&D\delta \delta A,\quad   D\bar\delta\delta A,\quad   D\bar\delta \bar\delta A,\quad   
\Delta\delta\delta A,\quad   \Delta\bar\delta\delta A,\quad   \Delta\bar\delta \bar\delta A,\nonumber\\
&D\bar\delta\bar\delta\bar C,\quad   \Delta\bar\delta\bar\delta \bar C,\quad   
\delta \bar\delta\bar\delta \bar C.
\end{align}
These computations require the evaluation of third-order commutation relations. After elimination of first and second order derivatives, we obtain algebraic relations for $A$ and $C$.
\begin{align}
A &=\frac{\mu}{\rho} B
+\frac{\bar\pi}{\bar\rho}C
-\frac{\bar\tau}{\bar\rho}\bar C, \tag{VII.4a}\label{A}\\
 \tau C &=\pi \bar C.\label{C} \tag{VII.4b}
\end{align}
We note that for $\tau=0$ (hence $\pi=0$) there is no algebraic relation for $C$, while  Eqn.(\ref{A}) is valid for both $\tau=0$ and $\tau\ne0$ cases.  
For $\tau$ and $\pi$ nonzero, using Eqn.(\ref{C}) and the identity $\tau\bar\tau=\pi\bar\pi$, it can be seen that  $\frac{\bar\pi}{\bar\rho}C
-\frac{\bar\tau}{\bar\rho}\bar C =0$. Then Eqn.(\ref{A}) reduces to 
$A=\frac{\mu}{\rho} B$
both for $\tau=0$ and $\tau\ne0$ cases.

For $\tau\ne 0$, it has been checked that all second-order equations for $A$, $B$, $C$, $\bar C$ and all first-order equations for $A$ and $C$ are satisfied.  The complete set of first-order equations for $B$ and $\bar C$ is given below.
\begin{eqnarray*}
&&DB=0,\\
&&\delta B=  (\bar\alpha +\beta -\tau) B -(\rho-\bar\rho)\bar C,\\
&&\bar\delta B =(\alpha +\bar\beta -\bar\tau)-(\bar\rho-\rho)C,\\
&&\Delta B=\left[  
(\gamma+\bar\gamma)
-\frac{\Psi_2\bar\tau +{\bar\Psi}_2 \pi+2\tau\bar\tau (\bar\tau+\pi)}
{\rho\bar\tau+\bar\rho\pi}
+(\rho-\bar\rho)\frac{\mu}{\rho}
\frac{\rho\bar\tau-\bar\rho\pi}{\rho\bar\tau+\bar\rho\pi}
\right ]B,\\
&&\\
&&D\bar C=-\rho \bar C-(\tau+\bar\pi)B,\\
&&\delta\bar C=-(\bar \alpha-\beta )\bar C,\\
&&\bar\delta \bar C=(\alpha-\bar\beta)\bar C +
(\rho\bar\tau+\bar\rho\pi)^{-1}
\left[ \frac{\pi}{\tau}\left(
\Psi_2\bar\rho-{\bar\Psi}_2\rho-2\rho^2 \bar\mu+2\bar\rho^2 \mu\right) 
-(\bar\tau+\pi)(\rho\bar\tau-\bar\rho\pi)\right]
\bar C,\\
&&\Delta\bar C=(\gamma-\bar\gamma+\bar\mu)\bar C
+\frac{\mu} {\rho}(\tau+\bar\pi)B.
\end{eqnarray*}
As $A$ and $C$  are algebraically determined and all derivatives of $B$ and $\bar C$  are determined, the freedom in the system consists of two arbitrary constants. Thus, we can state the following result.

\vskip 0.2cm\noindent
{\bf Proposition VII.1.} {\it \textcolor{red}{The twisting vacuum Type D metrics, whose principal null directions form a non-integrable distribution, admit two independent Killing vector fields.}}
\vskip 0.2cm\noindent
\textcolor{red}{In the case $\tau=0$ (and hence, since $\tau\bar{\tau}=\pi\bar{\pi}$, $\pi=0$), as $A$ is algebraically determined and only $\bar{\delta}\bar{C}$ remains undetermined, the system admits at most four independent Killing vector fields. The integrability conditions for $\tau=\pi=0$ are then evaluated and yield a system in involution with four arbitrary constants.}
\begin{align}
A &=\frac{\mu}{\rho}B,\quad DB=0,\quad \Delta B=0,\quad 
\delta B=-(\rho-\bar\rho)\bar C,\quad 
\bar \delta B=(\rho-\bar\rho) C, \tag{VII.5a}\\
DC &=-\bar\rho  C,\quad   \Delta C=\bar \mu C,\quad
\bar\delta C=-2\alpha C,\tag{VII.5b}\\
D\bar C &=-\rho  \bar C,\quad 
\Delta \bar C= \mu \bar C,\quad
\delta \bar C=-2\bar \alpha \bar C,\tag{VII.5c}\\
\delta C +\bar\delta\bar C &=-2 \bar \alpha C -2\alpha \bar C, \tag{VII.5d}\\
\bar \delta\bar\delta\bar C &-2\bar\delta(\alpha \bar C)
+(\rho-\bar\rho)^{-1}\left[
-\frac{\Psi_2 \bar\rho^2}{\rho}
+\frac{\bar\Psi_2\rho^2 }{\bar\rho}
\right ]C=0.\tag{VII.5e}
\end{align}
\textcolor{red}{Since the condition $\tau=0$ uniquely identifies the NUT solution among Type $D$ vacuum metrics, the number of independent Killing vector fields serves as a characterization of the NUT metric.}
\vskip 0.2cm\noindent
{\bf Proposition VII.2.} {\it \textcolor{red}{The NUT solution is the unique Type D vacuum metric admitting four dimensional Lie algebra of Killing vector fields.}}
\vskip 0.2cm\noindent
In the following, we \textcolor{red}{consider} the case $\tau=\pi=0$ and show that \textcolor{red}{ the spacetime admits four independent Killing vector fields forming a Lie algebra $u(1) \oplus su(2)$, corresponding to the Lie group $U(1) \times SU(2)$, as presented in \cite{yohannes,nozawa}.} We first show that the Killing vector with $C=0$ commutes with any Killing vector with $C\neq 0$.

\vskip 0.2cm\noindent
{\bf Proposition VII.3.} {\it 
Let $X^\circ = A^{\circ} D + B^{\circ} \Delta$, and $X = A D + B \Delta + C \delta + \bar{C} \bar{\delta}$ with $C\ne 0$ be Killing vectors of a Type $D$ vacuum metric with $\tau=\pi=0$.  Then $[X^\circ,X]=0$.}
\vskip 0.2cm\noindent
{\it Proof.} We replace $X^\circ$ and $X$ in the commutation relation, and use $DB^\circ=DB=0$  and $\Delta B^\circ=\Delta B=0$ to obtain 
\begin{align*}
[X^\circ,X]&=\left(A^{\circ}D+B^{\circ}\Delta\right) \left(AD+B\Delta+C\delta+\bar{C}\bar{\delta}\right)-\left(AD+B\Delta+C\delta+\bar{C}\bar{\delta}\right)\left(A^{\circ}D+B^{\circ}\Delta\right), \nonumber \\
    &=A^{\circ}(DA)D+A^{\circ}(DC)\delta+A^{\circ}(D\bar{C})\bar{\delta} + B^{\circ}(\Delta A)D + B^{\circ}(\Delta C)\delta + B^{\circ}(\Delta \bar{C})\bar{\delta} \nonumber \\
    &\quad - A(D A^{\circ})D- B (\Delta A^{\circ})D
-C  (\delta A^{\circ})D- C (\delta B^{\circ})\Delta 
-\bar C  (\bar\delta A^{\circ})D- \bar C (\bar\delta B^{\circ})\Delta\\
&=\left[  A^{\circ}(DA)
+ B^{\circ}(\Delta A)
 - A(D A^{\circ})
 - B (\Delta A^{\circ})
-C  (\delta A^{\circ})
-\bar C  (\bar\delta A^{\circ})\right]D\\
&\quad +\left[ - C (\delta B^{\circ})
- \bar C (\bar\delta B^{\circ})
\right]\Delta
+\left[ 
A^{\circ}(DC)+ B^{\circ}(\Delta C) 
\right ]\delta
+\left[A^{\circ}(D\bar{C}) 
+ B^{\circ}(\Delta \bar{C})
\right]\bar{\delta} 
\end{align*}
Note that as $X^\circ$ has no component along $\delta$ and $\bar\delta$, $\delta B^\circ=\bar\delta B^\circ=0$. Furthermore, as $A^\circ=(\mu/\rho) B^\circ$ and $\delta\mu=\delta\rho=\bar\delta\mu=\bar\delta\rho=0$, it follows that $\delta A^\circ=\bar\delta A^\circ=0$. Moreover, the commutation relation reduces to  
\begin{align*}
[X^\circ,X]&=\left[  A^{\circ}(DA) + B^{\circ}(\Delta A)  - A(D A^{\circ})  - B (\Delta A^{\circ}) \right]D +\left[ A^{\circ}(DC)+ B^{\circ}(\Delta C) \right ]\delta+\left[A^{\circ}(D\bar{C}) + B^{\circ}(\Delta \bar{C})\right]\bar{\delta}. 
\end{align*}
Replacing $A^\circ$, $DC$, $\Delta C$ and their conjugates, it is easy to see that the coefficients of $\delta$ and $\bar\delta$ vanish, and we are left with,
\begin{align*}
[X^\circ,X]&=\left[  A^{\circ}(DA) + B^{\circ}(\Delta A)  - A(D A^{\circ})  - B (\Delta A^{\circ}) \right]D . 
\end{align*}
Finally, using the expressions of $D(\mu/\rho)$ and $\Delta(\mu/\rho)$
one can see that the coefficient of $D$ also vanishes, hence $X^\circ$ and $X$ commute.
\hfill\qed

\vskip 0.2cm
\textcolor{red}{
We now show that the symmetry algebra is $su(2)$. 
According to the classification of three-dimensional Lie algebras~\cite{bowers2005classification}, 
the Lie algebra $su(2)$ is characterized by the property that for any $X, Y \in su(2)$, 
the set $\{X, Y, Z = [X, Y]\}$ is linearly independent.
}
\vskip 0.2cm\noindent
{\bf Lemma.} {\it For any $X$, $Y$ in $su(2)$, the set $\{X,Y,[X,Y]\}$ is linearly independent.}
\vskip 0.2cm\noindent
{\it Proof.}
For
$$X=A_xD+B_x \Delta +C_x \delta +\bar C_x \bar \delta, \quad Y= A_y D+B_y  \Delta +C_y \delta + \bar C_y \bar \delta,$$
their commutator is evaluated as
$$Z=A_z D +B_z \Delta +C_z \delta +\bar C_z \bar\delta.$$
If they are linearly independent, the matrix below should have rank 3.
$$M=\begin{pmatrix}
A_x & B_x & C_x & \bar C_x\\ 
A_y & B_y & C_y & \bar C_y\\
A_z & B_z & C_z & \bar C_z\\
\end{pmatrix}.
$$
In fact, the determinant of the sub-matrix consisting of the last $3$ columns is 

\begin{eqnarray*}
det(M)&=&  \delta\bar C_x     C_y   ( - B_x C_y   + B_y C_x )
+ \delta\bar C_y     C_x   (   B_x C_y   - B_y C_x  )
 + \delta     C_x     C_y   (   B_x \bar C_y - B_y \bar C_x)
+ \delta     C_y     C_x   ( - B_x \bar C_y + B_y \bar C_x)\\
&&  + \bar\delta\bar C_x \bar C_y( - B_x C_y + B_y C_x)
 + \bar\delta\bar C_y \bar C_x(   B_x C_y - B_y C_x)
 + \bar\delta C_x     \bar C_y(   B_x \bar C_y  - B_y \bar C_x )
  + \bar\delta C_y     \bar C_x( -B_x \bar C_y  + B_y \bar C_x )\\
&& +2 \alpha    ( - B_x \bar C_x \bar C_y C_y  + B_x \bar C_y^2 C_x          + B_y \bar C_x^2 C_y    - B_y \bar C_x \bar C_y C_x )\\
&& +2\bar\alpha ( - B_x \bar C_x      C_y^2    + B_x \bar C_y C_x C_y        + B_y \bar C_x C_x C_y  - B_y \bar C_y C_x^2 )\\
&& -(\rho-\bar\rho)    ( \bar C_x C_y-\bar C_y C_x)^2.
\end{eqnarray*}
In order to prove that this expression is nonzero, we proceed by proving the contrapositive.  Assume $det(M)=0$, then all its derivatives should be zero.  
In fact although $D(det(M))$ and $\Delta(det(M))$ are zero it turns out that $\delta(det(M))$  and $\bar\delta(det(M))$ are nonzero, proving that the set $\{X,Y,Z\}$ is linearly independent.\hfill \qed
\vskip 0.2cm\noindent
It follows that, 
\vskip 0.2cm\noindent
{\bf Proposition VII.4.} {\it 
The  symmetry  algebra of the NUT metric  is \textcolor{red}{$u(1) \oplus su(2)$}.}

\section{Non-twisting metrics} \label{sec4}
In this section, we provide a brief exposition of the consequences of the assumption $\rho = \bar{\rho}$ for vacuum Type $D$ space-times, where $\Psi_2$ is the only nonzero curvature component. Under these conditions, we have $\kappa = \nu = \sigma = \lambda = 0$, and since $\rho = \bar{\rho}$, one can use $SL(2,\mathbb{C})$ transformations to simultaneously set $\epsilon = 0$ and $\tau - \bar{\alpha} - \beta = 0$. We thus consider vacuum metrics for which $\Psi_2$ is the only nonzero component of the Weyl tensor, and the spin coefficients satisfy
\[ \rho=\bar\rho,\quad 
\kappa = \nu = \sigma = \lambda =0,\quad  \epsilon = 0, \quad 
\beta = \tau - \bar{\alpha}, \quad 
\pi = -\bar{\tau}.
\]
We will show that $\mu$ and $\Psi_2$ are also real, and, unlike in the twisting case, the condition $\tau + \bar{\pi} = 0$ does not imply $\tau = 0$. As $D\Psi_2=3\rho\Psi_2$ and $\Delta\Psi_2=-3\mu\Psi_2$, and $\Psi_2\ne 0$,  their commutator gives
$\Delta\rho+D\mu=(\gamma+\bar\gamma)\rho$.
Replacing  
\begin{eqnarray*}
&&\Delta\rho=
\bar\delta\tau
-2\alpha\tau+(\gamma+\bar\gamma-\bar\mu)\rho
-\Psi_2,\\
&&D\mu=-\delta\bar\tau+2\bar\alpha\tau +\mu\rho+\Psi_2,\\
&&\bar\delta\tau-\delta\bar\tau=
2\alpha\tau-2\bar\alpha\bar\tau-(\mu-\bar\mu)\rho+\Psi_2-{\bar\Psi}_2,
\end{eqnarray*}
in the commutation relation, we obtain $\Psi_2={\bar\Psi}_2$.  Then $\Delta\Psi_2=\Delta{\overline\Psi}_2$ implies  $\mu=\bar\mu$.
The system of NP equations is given below. 
\begin{align*}
&D\Psi_2=3\rho \Psi_2, &  
&\Delta\Psi_2= -3\mu \Psi_2,& \\       
&\delta \Psi_2=3\tau\Psi_2,&
&\bar\delta\Psi_2= 3\bar\tau \Psi_2, &  \\   
&D\rho= \rho^2,&
&\Delta\rho=\delta{\bar\tau}
-2\bar\alpha\bar\tau
+(\gamma+\bar\gamma-\mu)\rho
-\Psi_2,&\\
&\delta\rho=\tau\rho,&
&\bar\delta\rho=\bar\tau\rho,&
\\
&D\mu= -\delta\bar\tau
+2\bar\alpha\bar\tau+\mu\rho+\Psi_2,&  
&\Delta\mu=-(\gamma+\bar\gamma+\mu)\mu& \\   
&\delta\mu=-\tau\mu, & 
 &\overline{\delta}\mu=-\bar\tau\mu, & \\
&D\tau= 0,&  
&\Delta\tau=(\gamma-\bar\gamma)\tau,&  \\  
&\delta\tau=-2(\bar\alpha-\beta)\tau, &  &\overline{\delta}\tau=\delta\bar\tau+
2\alpha\tau-2\bar\alpha\bar\tau,& \\
 &D\alpha=\rho(\alpha-\bar\tau),&
 &\Delta \alpha =\bar\delta\gamma
 -\alpha(\gamma-\bar\gamma+\mu),&\\
 &\delta\alpha=-\bar\delta\bar\alpha +\delta\bar\tau
+ 4\alpha \bar\alpha  - \alpha \tau 
- 3\bar\alpha \bar\tau
+ \mu\rho + \tau\bar\tau -\Psi_2, 
& &&\\
&D\gamma=-\tau\bar\tau+\Psi_2,&
&\delta\gamma=-\delta\bar\gamma+2\mu\tau.&
&&
\end{align*}
The system for $\tau=0$ is integrable. For $\tau\ne 0$, there is a single integrability condition given as
$$\delta\delta\bar\tau
-2\delta(\bar\alpha\bar\tau)
-\tau(\delta\bar\tau-2\bar\alpha\bar\tau-2\mu\rho+2\Psi_2)=0.
$$
Thus, for $\rho=\bar\rho$ there are solutions with $\bar\tau+\pi=0$, but  $\tau\ne 0$, that correspond to  Type IIA metrics given by \cite{Kinnersley:1969zza}.

\section{Commutation relations for the metric components}\label{sec9}

Demonstrating that the set of NP equations \textcolor{red}{constitutes} an integrable system depending on a finite number of arbitrary constants means that the initial assumptions determine both the connection and the curvature up to these constants. The reconstruction of a metric with the given connection and curvature is equivalent to expressing the directional derivative operators, i.e, the moving frame, with respect to a commuting set of operators. \textcolor{red}{In the present context,} the $2$-dimensional sub-manifold whose tangent bundle is spanned by $m$ and $\bar m$ is a space of constant curvature. Accordingly, one retains the freedom of coordinate transformations in this $2$-dimensional sub-manifold. Consequently, we will express the derivative operators in terms of operators $D$, $\Delta_0$, $\delta_0$ and $\bar\delta_0$  such that $(\delta_0,\bar\delta_0)$ is the only non-commuting pair. The coefficients of the expansion in terms of these operators turn out to be analogues of the metric functions.  In analogy with the existing notation in the literature, we define,
\begin{equation}
    \Delta=UD+\Delta_0, \quad \delta=\bar\rho\delta_0. \nonumber
\end{equation}
where $\Delta_0$ and $\delta_0$ are differential operators commuting with $D$, i.e.,
$$ \Delta_0 D- D\Delta_0=0,\quad    \delta_0 D- D\delta_0=0.$$

\vskip 0.2cm\noindent {\bf Commutator of $D$ and $\Delta$:} 
The commutator of $D$ and $\Delta$ gives
$$
  (UD+\Delta_0)D-D(UD+\Delta_0)= 
  (\Delta_0 D- D\Delta_0)-(DU)D= (\gamma+\bar\gamma)D,
$$
since $D$  and $\Delta_0$ commute, we get
$DU=-(\gamma+\bar\gamma)$.
Using the definition of $\gamma$, we  can solve $U$ as
$$U=-\textstyle{\frac{1}{2}}\left (\rho\Psi_{20}+\bar\rho\bar{\Psi}_{20}     \right)+U_0,\quad DU_0=0.$$

\vskip 0.2cm\noindent {\bf  Commutator of $D$ and $\delta$:} 
The commutation relation of $D$ and $\delta$ 
$$
(\bar\rho \delta_0)D-D(\bar\rho\delta_0)= -\bar\rho (\bar\rho\delta_0)
$$
is identically satisfied.
\vskip 0.2cm\noindent {\bf  Commutator of $\Delta$ and $\delta$:} 
At the next step, we evaluate the commutator of $\Delta$ and $\delta$, using the fact that $\delta_0$ and $\Delta_0$  commute with $D$ along with the expression for $\Delta \bar{\rho}$.
\begin{eqnarray*}
  (\bar\rho\delta_0)(UD+\Delta_0)-(\Delta\bar\rho)\delta_0  -\bar\rho(UD+\Delta_0)\delta_0&=&(\mu-\gamma+\bar\gamma)(\bar\rho\delta_0),\\ 
  \bar\rho(\delta_0U)D
  +\bar\rho(\delta_0\Delta_0-\Delta_0\delta_0)
-(-\bar\rho\mu+(\gamma+\bar\gamma)\bar\rho-{\overline \Psi}_2)\delta_0
&=&
(\mu-\gamma+\bar\gamma)(\bar\rho\delta_0), \\ 
  \bar\rho(\delta_0U)D
  +\bar\rho(\delta_0\Delta_0-\Delta_0\delta_0)
&=&
(2\bar\gamma-{\overline \Psi}_2)(\bar\rho\delta_0)
\end{eqnarray*}
As $\gamma=\frac{1}{2}\frac{\Psi_2}{\rho}$, the right-hand side vanishes. Also as the commutator of $\delta_0$ and $\Delta_0$  have no $D$ component, it follows that
$$\delta_0U=0,\quad \delta_0\Delta_0-\Delta_0\delta_0=0. $$
\vskip 0.2cm\noindent {\bf  Commutator of $\delta$ and $\bar\delta$:} 
The last commutation relation is given by
\begin{eqnarray}
(\rho\bar\delta_0)(\bar\rho\delta_0)- (\bar\rho\delta_0)(\rho\bar\delta_0)&=&(\bar\mu-\mu)D+(\bar\rho-\rho)(UD+\Delta_0)
+2(\rho\alpha_0)(\bar\rho\delta_0)
-2(\bar\rho\bar\alpha_0)(\rho\bar\delta_0).\nonumber
\end{eqnarray}  
Using $\delta\rho=\delta\bar\rho$ and dividing by $\rho\bar\rho$, we obtain
\begin{eqnarray}
\bar\delta_0 \delta_0- \delta_0\bar\delta_0
&=&(\rho\bar\rho)^{-1}\left[(\bar\mu-\mu)+(\bar\rho-\rho)U\right]D
+\frac{\bar\rho-\rho}{\rho\bar\rho}\Delta_0
+2\alpha_0\delta_0
-2\bar\alpha_0\delta_0.\nonumber
\end{eqnarray}  
We evaluate the term in the bracket using  the expression of $\mu$ 
\begin{eqnarray}
(\bar\mu-\mu)+(\bar\rho-\rho)U&=&
-\textstyle\frac{1}{2}\left(  \frac{\Psi_2}{\rho}- \frac{\overline{\Psi}_2}{\bar\rho}\right)+(\bar\rho-\rho)
\left[ -\textstyle\frac{1}{2}( \frac{\Psi_2}{\rho^2}
+ \frac{\overline{\Psi}_2}{\bar\rho^2})+U_0            \right], \nonumber\\
&=&
-\textstyle\frac{1}{2}  \frac{\Psi_2}{\rho}
+\textstyle\frac{1}{2}  \frac{\overline{\Psi}_2}{\bar\rho}
-\textstyle\frac{1}{2}\bar\rho\frac{\Psi_2}{\rho^2}
-\textstyle\frac{1}{2}    \frac{\overline{\Psi}_2}{\bar\rho}
+\textstyle\frac{1}{2}   \frac{\Psi_2}{\rho}
+\textstyle\frac{1}{2}\rho\frac{\overline{\Psi}_2}{\bar\rho^2}
+(\bar\rho-\rho)U_0,\nonumber\\
&=&
-\textstyle\frac{1}{2}\bar\rho\frac{\Psi_2}{\rho^2}
-\textstyle\frac{1}{2}\rho\frac{\overline{\Psi}_2}{\bar\rho^2}
+(\bar\rho-\rho)U_0. \nonumber
\end{eqnarray}
and substitute in the commutation relation to obtain
\begin{eqnarray}
\bar\delta_0\delta_0- \delta_0\bar\delta_0&=&
(\rho\bar\rho)^{-1}
\left[    
-\textstyle\frac{1}{2}\bar\rho\frac{\Psi_2}{\rho^2}
+\textstyle\frac{1}{2}\rho\frac{\bar\Psi_2}{\bar\rho^2}
+(\bar\rho-\rho)U_0
\right]D
+(\rho\bar\rho)^{-1}
(\bar\rho-\rho)\Delta_0
+2\alpha_0 \delta_0
-2\bar\alpha_0  \delta_0,\nonumber\\
&=&
\left[    
-\textstyle\frac{1}{2}\Psi_{20}
+\textstyle\frac{1}{2}\bar\Psi_{20}
+(\rho\bar\rho)^{-1}(\bar\rho-\rho)U_0
\right]D
+(\rho\bar\rho)^{-1}
(\bar\rho-\rho)\Delta_0
+2\alpha_0 \delta_0
-2\bar\alpha_0  \delta_0.\nonumber
\end{eqnarray}  
Recall that
$    \frac{\bar\rho-\rho}{\rho\bar\rho}=\frac{1}{\rho}-\frac{1}{\bar\rho}=-2i\rho_0 $,
hence
\begin{equation}
    \bar\delta_0\delta_0- \delta_0\bar\delta_0=
\left[    
-\textstyle\frac{1}{2}\Psi_{20}
+\textstyle\frac{1}{2}\bar\Psi_{20}
-2i\rho_0 U_0
\right]D
-2i\rho_0 \Delta_0
+2\alpha_0 \delta_0
-2\bar\alpha_0  \delta_0. \nonumber
\end{equation}
As the coefficient of $D$ should be zero, we conclude that
$$-\textstyle\frac{1}{2}\Psi_{20}
+\textstyle\frac{1}{2}\bar\Psi_{20}
-2i\rho_0 U_0=0.$$
Thus, the final commutation relation simplifies to
\begin{equation}
\bar\delta_0\delta_0- \delta_0\bar\delta_0=
-2i\rho_0 \Delta_0
+2\alpha_0 \delta_0
-2\bar\alpha_0  \delta_0.  \nonumber   
\end{equation}

\begin{acknowledgments}
This work is funded by TUBITAK 1001 Program, Grant Number 123R114.
\end{acknowledgments}

\section*{Statements and Declarations}
\noindent \textbf{Funding:} This work is funded by TUBITAK 1001 Program, Grant Number 123R114.

\vspace{0.2cm}

\noindent \textbf{Competing Interests:} The authors have no relevant financial or non-financial interests to disclose.

\vspace{0.2cm}

\noindent \textbf{Author Contributions:} All authors contributed to the study conception and design. All authors read and approved the final manuscript.

\appendix
\section{The Newman-Unti-Tamburino (NUT) solution in local coordinates} \label{appnut}
Following \cite{Newman:1963yy}, in local coordinates $x^\mu=(t,r,x,y), (\mu=1,2,3,4)$, the covariant form of the NUT metric is given as
\begin{eqnarray}
g_{\mu \nu} = \left[
\begin{array}{cccc}
-2U  & 1 &  2A^{\circ}U & 2B^{\circ}U \\
1 & 0 &  -A^{\circ} & -B^{\circ}  \\
2A^{\circ}U &  -A^{\circ} &  -\left(R^2+2A^{\circ^2}U\right) & -2A^{\circ}B^{\circ}U  \\
2B^{\circ}U & -B^{\circ} & -2A^{\circ}B^{\circ}U &  -\left(R^2+2B^{\circ^2}U\right) \nonumber  
\end{array}
\right],
\end{eqnarray}
and the inverse metric is
\begin{eqnarray}
g^{\mu \nu} = \left[
\begin{array}{cccc}
- \left(A^{\circ^{2}}+B^{\circ^{2}}\right)R^{-2}  & 1 & -A^{\circ}R^{-2} & -B^{\circ}R^{-2} \\ 1 & 2U & 0 & 0 \\ -A^{\circ}R^{-2} & 0 & -R^{-2} & 0 \\ -B^{\circ}R^{-2} & 0 & 0 & -R^{-2}  \nonumber
\end{array}
\right],
\end{eqnarray}
where 
\begin{eqnarray}
    U &=& -\mu^{\circ}+\rho \overline{\rho} \left( r \Psi_{2x}^{\circ} + 2 \mu^{\circ} |\rho^{\circ}|^2\right), \quad
    A^{\circ} = \frac{|\rho^{\circ}|x^4}{\sqrt{2}p}, \quad 
    B^{\circ} = -\frac{|\rho^{\circ}|x^3}{\sqrt{2}p}, 
    \quad
   R^{-2} = 2 \rho \overline{\rho}p^2 
    = \left(-Det \ g^{\mu \nu}\right)^{1/2}, \nonumber
\end{eqnarray}
and 
\begin{eqnarray}    
    \rho &=&-\left(r+\rho^{\circ}\right)^{-1}, \ \  \rho \overline{\rho}=\frac{1}{r^2 + |\rho^{\circ}|^2}, \quad
    2 \mu^{\circ} = 1,0,-1, \ \ \ \ \ \  \   \sqrt{2} p=1+{\textstyle\frac{1}{2}}\mu^\circ (x^2+y^2). \nonumber 
\end{eqnarray}
Here, $\rho^{\circ}$ is pure imaginary and $\Psi_{2x}^{\circ}$ is a real constant.
The null tetrad can be chosen as 
\begin{eqnarray}
    l^{\mu} &=& \delta^{2}_{\mu}, \quad
    n^{\mu}= U \delta_{2}^{\mu} + \delta_1^{\mu}, \quad
    m^{\mu}= \overline{\rho}
(\xi^{\circ^1}\delta_1^{\mu}
+\xi^{\circ^3}\delta_3^{\mu}
+\xi^{\circ^4}\delta_4^{\mu}), \nonumber
\end{eqnarray}
where 
$$\xi^{\circ^1}=(\bar{\rho}p R^3)^{-1}(A^{\circ}+iB^{\circ}),\quad
\xi^{\circ^3}=-(\bar{\rho}p R^3)^{-1},\quad 
\xi^{\circ^4}=-i(\bar{\rho}p R^3)^{-1}. $$
Thus the covariant form of the null tetrad is 
\begin{eqnarray}
     l_{\mu}&=&\delta_{\mu}^1 - A^{\circ} \delta_{\mu}^3-B^{\circ} \delta_{\mu}^4 , \quad
    n_{\mu}= \delta_{\mu}^2+U\left[-\delta_{\mu}^1 +  A^{\circ} \delta_{\mu}^3+B^{\circ} \delta_{\mu}^4\right], \quad
    m_{\mu}= - \left(\delta_{\mu}^3 + i \delta_{\mu}^4\right)/2 \rho p. \nonumber
\end{eqnarray}

\section{NP Equations for Type $D$ Metrics}\label{appa}

\vskip 0.2cm\noindent
{\bf Commutation Relations.}
Commutation relations of the derivative operators  \cite{Stephani:2009exact} 
for the most general case are given below. 
\begin{align}
\Delta D-D\Delta&=
(\gamma+\bar{\gamma})D+(\epsilon+\bar{\epsilon})\Delta-(\tau+\bar{\pi})\bar{\delta}- (\bar{\tau}+\pi)\delta, \tag{B.1a} \label{B.1a}\\
\delta D-D\delta&=
(\bar{\alpha}+\beta-\bar{\pi})D+\kappa \Delta-\sigma \bar{\delta}-(\bar{\rho}+\epsilon-\bar{\epsilon})\delta,\tag{B.1b} \label{B.1b}\\
\delta \Delta-\Delta\delta&=
-\bar{\nu}D+(\tau-\bar{\alpha}-\beta)\Delta+\bar{\lambda}\bar{\delta}+(\mu-\gamma+\bar{\gamma})\delta, \tag{B.1c} \label{B.1c} \\
\bar{\delta}\delta-\delta\bar{\delta}&=
(\bar{\mu}-\mu)D+(\bar{\rho}-\rho)\Delta-(\bar{\alpha}-\beta)\bar{\delta}-(\bar{\beta}-\alpha)\delta. \tag{B.1d} \label{B.1d}
\end{align} 
From these equations, it follows that for Type $D$ metrics, where $D$ and $\Delta$ are principal null directions of the Weyl tensor (so that $\kappa = \nu = \sigma = \lambda = 0$), the pairs $\{D, \delta\}$ and $\{\Delta, \delta\}$ form integrable distributions. In addition, if $\tau + \bar{\pi} = 0$, the pair $\{D, \Delta\}$ also forms an integrable distribution. On the other hand, when $\rho \neq \bar{\rho}$ and $\mu \neq \bar{\mu}$, the distribution spanned by $\{\delta, \bar{\delta}\}$ is not integrable.

\vskip 0.2cm\noindent
{\bf Exterior Derivatives.}
The exterior derivatives of the basis $1$-forms for the general case \cite{bilgedaghan,Frolov:1998wf} 
are given below.
\begin{align}
dl&=
-(\epsilon+\bar{\epsilon})          ln
+(\alpha+\bar{\beta}-\bar{\tau})    lm
+(\bar{\alpha}+{\beta}-{\tau})      l\bar{m}
-\bar{\kappa}                       nm
-\kappa                             n\bar{m}
+(\rho-\bar{\rho})                  m\bar{m}, \tag{B.2a} \label{B.2a}\\
dn&=
-(\gamma+\bar{\gamma})              ln
+\nu                                lm
+\bar{\nu}                          l\bar{m}
+(\pi-\alpha-\bar{\beta})           nm
+(\bar{\pi}-\bar{\alpha}-{\beta})   n\bar{m}
+(\mu-\bar{\mu})                    m\bar{m},\tag{B.2b} \label{B.2b}\\
dm&=
-(\tau+\bar{\pi})                   ln
+(\bar{\mu}+\gamma-\bar{\gamma})    lm
+\bar{\lambda}                      l\bar{m}
+(\epsilon-\bar{\epsilon}-\rho)     nm
-\sigma                             n\bar{m}
-(\bar{\alpha}-\beta)               m\bar{m}.\tag{B.2c} \label{B.2c}
\end{align}
\vskip 0.2cm\noindent
{\bf Bianchi Identities and NP Equations.}
The set of NP equations and Bianchi identities  for  Type $D$ vacuum metrics,  with
$$\Psi_i=0,\quad i\ne 2,\quad \Phi_{ij}=0,\quad i,j=0,1,2,\quad \Lambda=0,$$
is given below.
\begin{align}
 D\Psi_2&=3\rho\Psi_2, \tag{B.3a} \label{B.3a}\\
\Delta\Psi_2&=-3\mu\Psi_2, \tag{B.3b} \label{B.3b}\\
\delta\Psi_2&=3\tau\Psi_2, \tag{B.3c} \label{B.3c}\\
\bar\delta\Psi_2&=-3\pi\Psi_2, \tag{B.3d} \label{B.3d}\\
& \nonumber \\
D\rho   &= \rho^2,       \tag{B.4a} \label{B.4a}       \\
\Delta\rho      &=\bar\delta\tau +(\bar\beta -\alpha -\bar\tau) \tau
+ (\gamma+\bar \gamma- \bar \mu)\rho -\Psi_2, \tag{B.4b} \label{B.4b}\\
\delta\rho      &= \rho(\bar\alpha+\beta)
                    +(\rho-\bar\rho)\tau, \tag{B.4c} \label{B.4c}\\
& \nonumber \\
D\mu  &=\delta\pi+ \bar\rho\mu+
\pi(\bar \pi -\bar\alpha+\beta)+\Psi_2, \tag{B.5a} \label{B.5a}\\
\Delta\mu      &= -(\mu+\gamma+\bar\gamma)\mu,\tag{B.5b} \label{B.5b} \\
\bar\delta\mu  &= -(\mu-\bar\mu)\pi-\mu(\alpha+\bar\beta),\tag{B.5c} \label{B.5c}  \\
& \nonumber \\
D\tau   &= (\tau+\bar\pi)\rho,  \tag{B.6a} \label{B.6a}  \\
\delta\tau      &= (\tau+\beta-\bar\alpha)\tau, \tag{B.6b} \label{B.6b} \\
\Delta \pi     &= -(\pi+\bar\tau+\gamma-\bar\gamma)\mu, \tag{B.6c} \label{B.6c} \\
\bar\delta\pi  &= -(\pi+\alpha-\bar\beta)\pi, \tag{B.6d} \label{B.6d} \\
& \nonumber \\
D\alpha &= \rho(\alpha+\pi), \tag{B.7a} \label{B.7a}   \\
D\beta  &= \bar\rho\beta,  \tag{B.7b} \label{B.7b}      \\
\Delta\alpha  &=\bar\delta\gamma +
(\bar\gamma-\bar\mu)\alpha+(\bar\beta-\bar\tau)\gamma, \tag{B.7c} \label{B.7c}\\
\Delta\beta  &= \delta\gamma  +
(\tau-\bar\alpha-\beta)\gamma+\mu\tau
-\beta(\gamma-\bar\gamma-\mu),\tag{B.7d} \label{B.7d}       \\
\delta\alpha  &= \bar\delta\beta + \mu\rho+\alpha\bar\alpha
                    +\beta\bar\beta-2\alpha\beta
                    +\gamma(\rho-\bar\rho)-\Psi_2,\tag{B.7e} \label{B.7e} \\
& \nonumber \\
D\gamma &= (\tau+\bar\pi)\alpha+             (\bar\tau+\pi)\beta+\tau\pi+\Psi_2, \tag{B.8} \label{B.8}
\end{align}
\vskip 0.2cm\noindent
{\bf Killing Equations.}
Let $X$ be a Killing vector parametrized as
\begin{equation}
    X=AD+B\Delta+C\delta+\bar C\bar\delta.\tag{B.9} \label{B.9}
\end{equation}
The Killing equation in NP formalism gives the following $10$ equations 
in the most general case \cite{basov2012problems}.
\begin{align}
&\left(D+\epsilon+\bar{\epsilon}\right)A + \left(\Delta-\gamma - \bar{\gamma}\right)B - \left(\tau- \bar{\pi}\right)C - \left(\bar{\tau}-\pi\right)\bar{C}=0, \tag{B.10a} \label{B.10a}\\
&\left(D+\epsilon+\bar{\epsilon}\right)B -C \kappa - \bar{C} \bar{\kappa}=0, \tag{B.10b} \label{B.10b}\\
& \left(D+\rho+\epsilon-\bar{\epsilon}\right)C-A \bar{\kappa} - \left(\bar{\delta}-\alpha-\bar{\beta}-\pi\right)B  + \bar{C} \bar{\sigma}=0, \tag{B.10c} \label{B.10c} \\
&  \left(D+\bar{\rho}+\bar{\epsilon}-\epsilon\right)\bar{C}-A \kappa - \left(\delta-\bar{\alpha}-\beta-\bar{\pi}\right)B + C \sigma =0,  \tag{B.10d} \label{B.10d}\\
&\left(\Delta+\gamma +\bar{\gamma}\right)A + C \bar{\nu} +\bar{C} \nu =0, \tag{B.10e} \label{B.10e}\\
& \left(\Delta+\gamma-\bar{\gamma}-\bar{\mu}\right)C -\left(\bar{\delta}+\bar{\tau}+\alpha+\bar{\beta}\right)A + B \nu - \bar{C}\lambda=0, \tag{B.10f} \label{B.10f} \\
& \left(\Delta+\bar{\gamma}-\gamma-\mu\right)\bar{C}-\left(\delta+\tau+\bar{\alpha}+\beta\right)A + B \bar{\nu}  - C\bar{\lambda}=0, \tag{B.10g} \label{B.10g} \\
& \left(\bar{\delta}+\alpha-\bar{\beta}\right)C-A \bar{\sigma} + B \lambda =0, \tag{B.10h} \label{B.10h} \\
&\left(\delta+\bar{\alpha}-\beta\right)\bar{C}- A \sigma +B \bar{\lambda} =0, \tag{B.10i} \label{B.10i}\\
 & \left(\delta-\bar{\alpha}+\beta\right)C + \left(\bar{\delta}-\alpha+\bar{\beta}\right)\bar{C}-A\left(\rho+\bar{\rho}\right) + B \left(\mu + \bar{\mu}\right)=0. \tag{B.10j} \label{B.10j}
\end{align}

\bibliographystyle{apsrev4-2}
\bibliography{Referanslar}

@article{Newman:1961qr,
  title={An approach to gravitational radiation by a method of spin coefficients},
  author={Newman, Ezra and Penrose, Roger},
  journal={Journal of Mathematical Physics},
  volume={3},
  number={3},
  pages={566--578},
  year={1962},
  publisher={American Institute of Physics}
}

@book{Stephani:2009exact,
  title={Exact solutions of Einstein's field equations},
  author={Stephani, Hans and Kramer, Dietrich and MacCallum, Malcolm and Hoenselaers, Cornelius and Herlt, Eduard},
  year={2009},
  publisher={Cambridge University Press}
}

@article{Carmeli:1975wq,
  title={Transformation laws of the Newman-Penrose field variables},
  author={Carmeli, M and Kaye, M},
  journal={Annals of Physics},
  volume={99},
  number={1},
  pages={188--195},
  year={1976},
  publisher={Elsevier}
}

@article{schwarz1984,
  title={The Riquier-Janet theory and its application to nonlinear evolution equations},
  author={Schwarz, Fritz},
  journal={Physica D: Nonlinear Phenomena},
  volume={11},
  number={1-2},
  pages={243--251},
  year={1984},
  publisher={Elsevier}
}

@article{ahsan,
  title={A Solution of Weyl-Lanczos Equations for Arbitrary Petrov Type D Vacuum Spacetimes},
  author={Ahsan, Zafar and Bilal, Mohd},
  journal={International Journal of Theoretical Physics},
  volume={49},
  pages={2713--2722},
  year={2010},
  publisher={Springer}
}

@article{brans1967computer,
  title={A Computer Program for the Nonnumerical Testing and Reduction of Sets of Algebraic Partial Differential Equations},
  author={Brans, Carl H},
  journal={Journal of the ACM (JACM)},
  volume={14},
  number={1},
  pages={45--62},
  year={1967},
  publisher={ACM New York, NY, USA}
}

@book{Stephani2003,
  title={``Part III: The various classes of algebraically special solutions" in Exact solutions of Einstein's field equations},
  author={Stephani, Hans and Kramer, Dietrich and MacCallum, Malcolm and Hoenselaers, Cornelius and Herlt, Eduard},
  year={2003},
  publisher={Cambridge University Press}
}

@article{czapor1982orthogonal,
  title={Orthogonal transitivity, invertibility and null geodesic separability in type D vacuum solutions of Einstein’s field equations with cosmological constant},
  author={Czapor, Stephen R and McLenaghan, RG},
  journal={Journal of Mathematical Physics},
  volume={23},
  number={11},
  pages={2159--2167},
  year={1982},
  publisher={American Institute of Physics}
}

@article{debever1981orthogonal,
  title={Orthogonal transitivity, invertibility, and null geodesic separability in type D electrovac solutions of Einstein’s field equations with cosmological constant},
  author={Debever, Robert and McLenaghan, RG},
  journal={Journal of Mathematical Physics},
  volume={22},
  number={8},
  pages={1711--1726},
  year={1981},
  publisher={American Institute of Physics}
}

@article{edgar2009petrov,
  title={Petrov D vacuum spaces revisited: identities and invariant classification},
  author={Edgar, S Brian and G{\'o}mez-Lobo, Alfonso Garc{\'\i}a-Parrado and Mart{\'\i}n-Garc{\'\i}a, Jos{\'e} M},
  journal={Classical and Quantum Gravity},
  volume={26},
  number={10},
  pages={105022},
  year={2009},
  publisher={IOP Publishing}
}

@article{Kinnersley:1969zza,
  title={Type D vacuum metrics},
  author={Kinnersley, William},
  journal={Journal of Mathematical Physics},
  volume={10},
  number={7},
  pages={1195--1203},
  year={1969},
  publisher={AIP Publishing}
}

@article{Newman:1963yy,
  title={Empty-space generalization of the Schwarzschild metric},
  author={Newman, Ezra and Tamburino, L and Unti, Theodore},
  journal={Journal of Mathematical Physics},
  volume={4},
  number={7},
  pages={915--923},
  year={1963},
  publisher={American Institute of Physics}
}

@article{bilgedaghan,
  title={Warped Product Metrics with Rank-1 Ricci Tensor},
  author={Bilge, AH and Daghan, D},
  journal={European Astronomical Society Publications Series},
  volume={30},
  pages={329--332},
  year={2008},
  publisher={EDP Sciences}
}

@article{maia,
  title={Cosmology of spin-2 fields},
  author={Maia, MD},
  journal={International Journal of Modern Physics A},
  volume={31},
  number={02n03},
  pages={1641010},
  year={2016},
  publisher={World Scientific}
}

@article{izumi2013analysis,
  title={An analysis of characteristics in nonlinear massive gravity},
  author={Izumi, Keisuke and Ong, Yen Chin},
  journal={Classical and Quantum Gravity},
  volume={30},
  number={18},
  pages={184008},
  year={2013},
  publisher={IOP Publishing}
}

@misc{ita2008proposal,
      author={Ita III, Eyo Eyo},
      year={2013},
      eprint={0806.3964},
      archivePrefix={arXiv},
      primaryClass={gr-qc},
      url={https://arxiv.org/abs/0806.3964}, 
}

@article{klainerman1999local,
  title={On local and global aspects of the Cauchy problem in general relativity},
  author={Klainerman, Sergiu and Nicol{\`o}, Francesco},
  journal={Classical and Quantum Gravity},
  volume={16},
  number={8},
  pages={R73},
  year={1999},
  publisher={IOP Publishing}
}

@article{lanczos1949lagrangian,
  title={Lagrangian multiplier and Riemannian spaces},
  author={Lanczos, Cornelius},
  journal={Reviews of Modern Physics},
  volume={21},
  number={3},
  pages={497},
  year={1949},
  publisher={APS}
}

@article{bampi1983third,
  title={Third-order tensor potentials for the Riemann and Weyl tensors},
  author={Bampi, Franco and Caviglia, Giacomo},
  journal={General Relativity and Gravitation},
  volume={15},
  pages={375--386},
  year={1983},
  publisher={Springer}
}

@article{dolan2003exterior,
  title={Exterior differential systems, Janet--Riquier theory and the Riemann--Lanczos problems in two, three, and four dimensions},
  author={Dolan, P and Gerber, A},
  journal={Journal of Mathematical Physics},
  volume={44},
  number={7},
  pages={3013--3034},
  year={2003},
  publisher={American Institute of Physics}
}

@book{Frolov:1998wf,
    editor = "Frolov, V. P. and Novikov, I. D.",
    title = "{Black hole physics: Basic concepts and new developments}",
    doi = "10.1007/978-94-011-5139-9",
    year = "1998",
    publisher={Springer}
}

@article{nozawa,
   title={An alternative to the Simon tensor},
   volume={38},
   ISSN={1361-6382},
   url={http://dx.doi.org/10.1088/1361-6382/ac0a87},
   DOI={10.1088/1361-6382/ac0a87},
   number={15},
   journal={Classical and Quantum Gravity},
   publisher={IOP Publishing},
   author={Nozawa, Masato},
   year={2021},
   month=jul, pages={155001} }

@misc{yohannes,
      author={Schiden Yohannes and Domenico Giulini},
      year={2021},
      eprint={2102.08496},
      archivePrefix={arXiv},
      primaryClass={math-ph},
      url={https://arxiv.org/abs/2102.08496}, 
}

@misc{bowers2005classification,
  author = {Bowers, Adam},
  year={2005},
  note = {\url{https://math.ucsd.edu/~abowers/downloads/survey/3d_Lie_alg_classify.pdf} [Accessed: 2025-09-10]},
  urldate = {2025-09-10}
}

@book{basov2012problems,
  title={Problems in the general theory of relativity and theory of group representations},
  author={Basov, Nikola Gennadievich},
  year={2012},
  publisher={Springer Science \& Business Media}
}

@article{Petrov:2000bs,
    author = "Petrov, A. Z.",
    title = "{The Classification of spaces defining gravitational fields}",
    doi = "10.1023/A:1001910908054",
    journal = "Gen. Rel. Grav.",
    volume = "32",
    pages = "1665",
    year = "2000"
}

@article{McIntosh_1994,
author = {C B G McIntosh and R Arianrhod and S T Wade and C Hoenselaers},
year = {1994},
month = {jun},
publisher = {},
volume = {11},
number = {6},
pages = {1555},
title = {Electric and magnetic Weyl tensors: classification and analysis},
journal = {Classical and Quantum Gravity},
}

@article{dinverno71,
    author = {d'Inverno, R. A. and Russell‐Clark, R. A.},
    title = {Classification of the Harrison Metrics},
    journal = {Journal of Mathematical Physics},
    volume = {12},
    number = {7},
    pages = {1258-1263},
    year = {1971},
    month = {07},
    issn = {0022-2488},
    doi = {10.1063/1.1665729},
    url = {https://doi.org/10.1063/1.1665729},
}

@article{Baak2023,
    author = "Baak, Sang-Shin and Datta, Satadal and Fischer, Uwe R.",
    title = "{Petrov classification of analogue spacetimes}",
    eprint = "2305.12771",
    archivePrefix = "arXiv",
    primaryClass = "gr-qc",
    doi = "10.1088/1361-6382/acf08e",
    journal = "Class. Quant. Grav.",
    volume = "40",
    number = "21",
    pages = "215001",
    year = "2023"
}

@article{Ferrando2002,
    author = "Ferrando, Joan Josep and Saez, Juan Antonio",
    title = "{On the classification of type D space-times}",
    eprint = "gr-qc/0212086",
    archivePrefix = "arXiv",
    doi = "10.1063/1.1640795",
    journal = "J. Math. Phys.",
    volume = "45",
    pages = "652--667",
    year = "2004"
}
\end{document}